\documentclass[a4paper,12pt]{article}

\usepackage[utf8x]{inputenc}
\usepackage{graphicx}
\usepackage{amsfonts}
\usepackage{amsmath}
\usepackage{amssymb}
\usepackage{cite}
\usepackage{hyperref}
\usepackage{bookmark}

\allowdisplaybreaks

\begin{document}

\date{}

\title{\textbf{An Observer-Based View of Euclidean Geometry}}
\author{\textbf{Newshaw Bahreyni}$^{1}$, \textbf{Carlo Cafaro}$^{2,3}$, \textbf{Leonardo Rossetti}$^{4,2}$ \\ $^{1}$Pomona College, Claremont, CA 91711, USA\\   
$^{2}$University at Albany-SUNY, Albany, NY 12222, USA\\  
$^{3}$SUNY Polytechnic Institute, Utica, NY 13502, USA\\
$^{4}$University of Camerino, I-62032 Camerino, Italy }

\maketitle

\begin{abstract}
Influence network of events is a view of the universe based on events that may be related to one another via influence. The network of events form a partially-ordered set which, when quantified consistently via a technique called chain projection, results in the emergence of spacetime and the
Minkowski metric as well as the Lorentz transformation through changing an observer from one frame to another. Interestingly, using this approach, the motion of a free electron as well as the Dirac equation can be described.  Indeed, the same approach can be employed to show how a discrete version of
some of the features of Euclidean geometry, including directions, dimensions, subspaces, Pythagorean theorem, and geometric shapes can emerge.

In this paper, after reviewing the essentials of the influence network formalism, we build on some of our previous works to further develop aspects of Euclidean geometry. Specifically, we present the emergence of geometric shapes, a discrete version of the Parallel postulate, the dot product, and
the outer (wedge product) in $2+1$ dimensions. Finally, we show that the scalar quantification of two concatenated orthogonal intervals exhibits features that are similar to those of the well-known concept of geometric product in geometric Clifford algebras.
\end{abstract}

\pagebreak
\tableofcontents


\section{Introduction}
Euclid's book \textquotedblleft the Elements\textquotedblright, which has been the foundation of Euclidean geometry, presents a list of definitions and postulates based on logic, common sense and the perceptions of the surrounding space as the basis for proving the rest of the theorems in flat space \cite{Euclid}.  Hence, the definitions are considered to be fundamental.  For example, Euclid defines angles, triangles and areas and then gives a proof for the Pythagorean theorem by rearranging areas of identical triangles.  Therefore angles, triangles and areas are treated to be fundamental and the Pythagorean theorem is proved based on these fundamental definitions which implies that the Pythagorean theorem is not fundamental itself.

As the study of the surrounding objects is not independent of studying the surrounding space, studying geometry has been tied to studying space since its early days.  The ideas about the nature of space viewed it to be infinite and fundamental \cite{Jammer}.  Later with Newton's laws space was believed to be independent of what happens in it until Einstein showed that space is shaped by what goes on in it \cite{Einstein}.  This questioned the absoluteness of space but only to the extent that scientists started looking for the shape of the universe based on the amount of matter and energy it contained; the absoluteness of space was still intact.  Einstein's Relativity theory gave rise to a new geometry to explain the properties of a curved space.  However, like the Euclidean geometry, in this new geometry the fundamental notions such as angles, geometrical shapes and areas were still considered to be fundamental and serve as the basis for the derivation of other properties.  

Here the universe is viewed as a partially-ordered set (poset) of events where events are defined as interactions between particles.  Thus a causal relation is formed as introduced and discussed \textbf{in \cite{KB1}\cite{KB2}\cite{Kquant}\cite{Kcontemp}\cite{Kelectron}\cite{KInfo1}\cite{KInfo2}} which results from probabilistic inferences \cite{KS}.  Partially-ordered sets are the same as directed acyclic graphs (DAG).  The causal influence in a partially-ordered set of events called \textit{causal sets} or \textit{causets} were introduced by Bombelli and collaborators and studied extensively by Sorkin, where the causal sets are usually embedded in a Minkowski geometry that exhibits Lorentz invariance \cite{Bombelli1}\cite{Bombelli2}\cite{Sorkin1}\cite{Sorkin2}.  This approach where a causal model is embedded in spacetime in an information-theoretic setting has been used to find connections between geometric and informational causality \cite{GV}.  Causal models have also been used to find new interpretations of quantum theory \cite{OB}.  The concept of information flow plays a key role in causal influences that are studied as an information-theoretic setting \cite{C1}\cite{C2}.  
Interestingly, in Ref. \cite{D} the geometry of spacetime was shown to emerge from abstract order lattices \cite{GB}\cite{DP} expressed in a quantitative way by means of the geometric Clifford algebra language \cite{DL}\cite{H}\cite{C3}\cite{C4}.

Geometric algebra which is Clifford's generalization of complex numbers and quaternion algebra to vectors in arbitrary dimensions is a formalism in which elements of any grade such as scalars, vectors, bivectors, and higher order multivectors can be added or multiplied together.  This result is shown in the geometric product of two vectors $\vec{a}$ and $\vec{b}$ as the sum of an inner product and an outer product 
\begin{equation*}
\vec{a}\vec{b} = \vec{a}.\vec{b} + \vec{a} \wedge \vec{b}.
\end{equation*}
In three dimensional space, $\vec{a}$ and $\vec{b}$ are three dimensional vectors and this  relation gives the usual results from the dot and the cross products \cite{DL}.  This algebra has been used to describe physical laws in different fields of physics.    

In our picture, instead of assuming events taking place in any space or time or assuming any properties for them, we take the poset to be fundamental and seek a consistent quantification such that it preserves the order by mapping events to real numbers.  Quantification of lattices which are a special case of posets has been studied by Knuth where instead of assuming sum and product rules, he has demonstrated how quantifications constrained by lattice symmetries result in constraint equations representing the sum and product rules \cite{Klattice1}\cite{Klattice2}\cite{Klattice3}.

We quantify this causal set of events using embedded observers represented by chains that are totally ordered sets of events.  In our previous work we showed how consistently quantifying the poset using a pair of embedded observers resulted in a discrete version of the Minkowski metric and the Lorentz transformations \cite{KB1}\cite{KB2}\cite{Kquant}\cite{KInfo2}. Kinematics also arises in this picture as a result of changing relationship between observer and object \cite{KW}.  The focus of this paper is to investigate the geometrical results from our previous work such as directionality, subspaces, and the Pythagorean theorem further and study quantifications with more than a pair of embedded observers.  We show how quantifying with more than two observers is related to higher than 1+1 dimensions.  In addition, we show how constraints from quantifying with a number of embedded observers results in some fundamental features of the Euclidean geometry such as the Parallel postulate, a discrete version of the dot and a discrete version of the outer (wedge) product in 2+1 dimensions.

In this paper we will review our previous work on the quantification technique of a poset using an embedded observer in section 2.  In section 3 we will review the coordination condition which constraints the quantification using a pair of embedded observers, the conditions for directionality, subspaces and the Pythagorean theorem.  We will then show in section 4 how quantifying with more than a pair of observers results in simplices and subspaces up to N+1 dimensions where $N\in {1,2,3,...,\infty}$.  In sections 5 and 6 we discuss quantification of the poset using a number of coordinated and collinear embedded observers, called a \textit{fence}, and present a derivation of a discrete version of the Parallel postulate and the dot product.  Finally, in section 7 we will extend the quantification from a set of collinear and coordinated observers (a fence) to multiple sets of fences, called a \textit{grid} and show how quantification for an orthogonal grid which is a special case of a grid leads to a discrete version of the outer (wedge) product in 2+1 dimensions.  In section 8, we present a summary of results along with our final remarks.

\section{Influence Network and Its Quantification}
We start by reviewing the definition of a partially-ordered set of events and it quantification using an embedded chain which we call an observer. 
\subsection{Partially-Ordered Set of Events and Chains}
A \textit{partially-ordered set} $\textit{P}$ is a set of elements with a binary ordering relation $\leq$, called \textit{inclusion}, such that for elements $a,b,c\in \textit{P}$ inclusion is transitive (if $x\leq y$ and $y\leq z$, then $x\leq z$), reflexive ($\forall x\in \textit{P},\: x\leq x$) and antisymmetric (if $x\leq y$ and $y\leq x$ then $x=y$) \cite{KBook}.  If elements of the poset are not related through inclusion, they are incomparable, denoted $a\ll b$, hence the name partially-ordered set of elements. 

If for two elements of the poset $a,b\in \textit{P}$ we have that $a<b$ and $a\neq b$ and there is not an element $x$ such that $a<x<b$, then $b$ covers $a$, denoted $a\prec b$.       
\textbf{In \cite{KB1}\cite{KB2}\cite{Kquant}\cite{Kcontemp}} we defined an \textit{event} as the boundary of an interaction where one event represents the act of influencing and another event represents the act of being influenced.  A \textit{partially-ordered set of events}, or a poset of events, is defined as a set of events $\Pi$ and a binary relation $\leq$ defined by influence such that this relation is transitive (if $x\leq y$ and $y\leq z$, then $x\leq z$), reflexive ($\forall x\in \Pi,\: x\leq x$) and antisymmetric (if $x\leq y$ and $y\leq x$ then $x=y$) \cite{KB1}\cite{KB2}\cite{Kquant}\cite{Kcontemp}. 

A \textit{chain} is defined to be a set of events such that for any given $x$ and $y$ in the chain, either $x\leq y$ or $y\leq x$ \cite{Birkhoff}.  Therefore a chain is totally ordered.  Given a poset $\Pi$ with event $x\in \Pi$ and chain $\textbf{P}\in \Pi$ such that $\exists p_{i}\in \textbf{P}$ where $x\leq p_{i}$ then the \textit{forward projection} of $x$ onto $\textbf{P}$, is given by the map (functional) $P: x \in \Pi \rightarrow p_{x}\in \textbf{P}$ such that $p_{x}= min \left\{p_{i}\in \textbf{P}|x\leq p_{i}\right\}$ and if $\exists p_{i}\in \textbf{P}$ where $x\geq p_{i}$ then the \textit{backward projection} of $x$ onto $\textbf{P}$, is given by the map (functional) $\overline{P}: x \in \Pi \rightarrow \overline{p}_{x}\in \textbf{P}$ such that $\overline{p}_{x}= max \left\{p_{i}\in \textbf{P}|x\geq p_{i}\right\}$ \cite{KB1}\cite{KB2}.

\begin{figure}[t]
\centering
\includegraphics[width=0.80%
\textwidth]{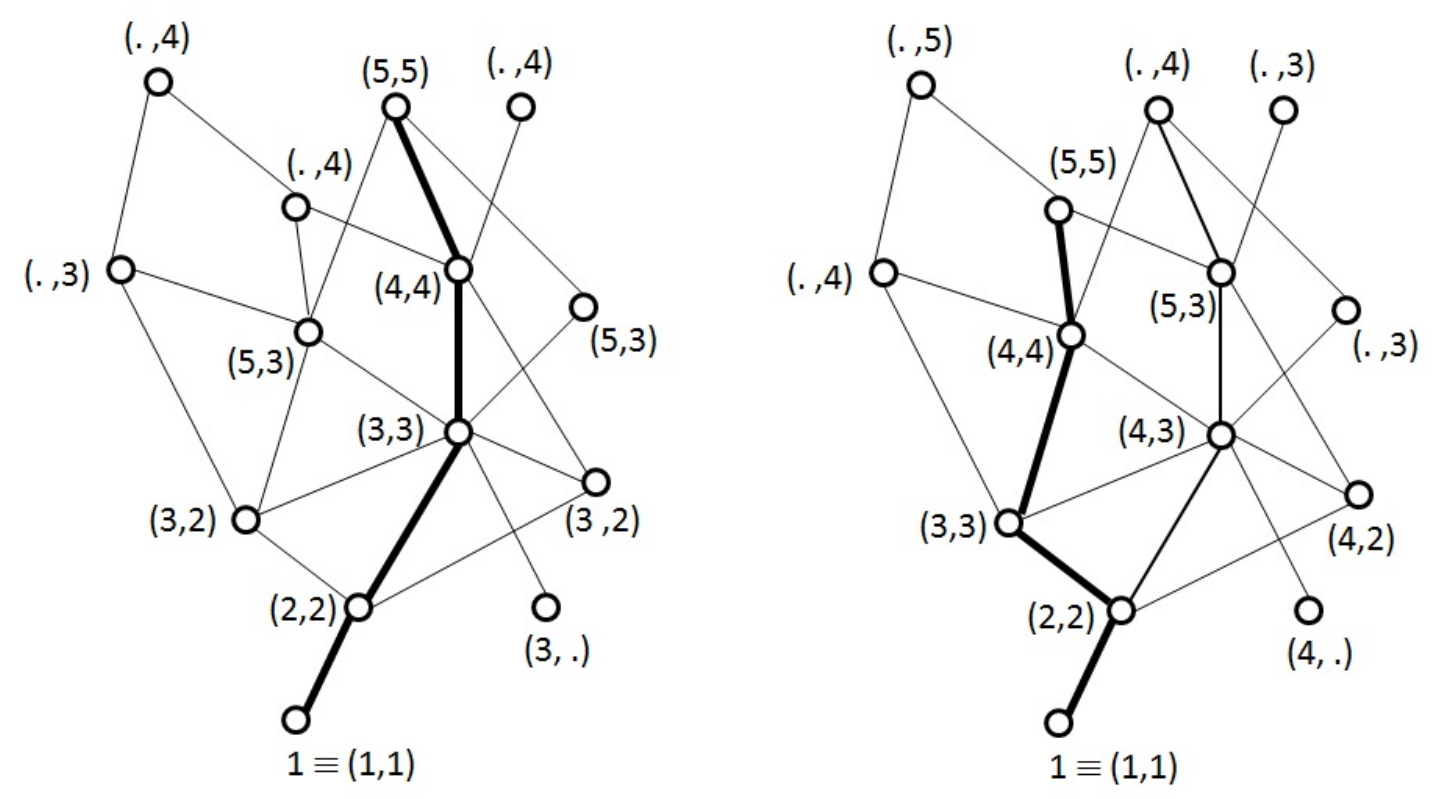}  
\caption{Quantification of a poset with respect to two different embedded
observers (chains). Embedded observers are totally ordered set of events
that are quantified by isotonic valuations of successive integers and the
rest of the poset is quantified with respect to the observers using the
forward and backward projections. Quantification of the poset will be
different according to each embedded observer shown in the two figures.}
\label{fig:Figure1}
\end{figure}

Quantification of a partially-ordered set of events is carried out through chain projection, as shown in Fig. \ref{fig:Figure1}.  Elements of a chain $\textbf{P}$ are quantified by defining a functional called \textit{monotonic valuation} that takes each element $p$ of the chain to a real number $v_{\textbf{P}}(p)$ such that if $x\leq y$ then $v_{\textbf{P}}(x)\leq v_{\textbf{P}}(y)$, where $v_{\textbf{P}}(x)$ and $v_{\textbf{P}}(y)$ are valuations assigned to $x$ and $y$ with respect to chain $\textbf{P}$, respectively.  Other events of the poset that are forward and backward projected onto the elements of the distinct chain will be quantified by the corresponding valuations of the two events onto which they are projected.  Note that the quantification of an element by chain projection is chain-dependent.  Furthermore, there are some subsets of events that cannot be quantified by any given chain.  However, these subsets are different for different chains.

An \textit{interval} is comprised of a pair of events that are quantified by the same chain or the same pair of chains.  There are different classes of intervals: chain-like intervals, antichain-like intervals, and projections-like intervals. 

An interval is called \textit{chain-like} iff both elements of its quantifying pair are of like sign, with $(0,0)$ as a degenerate case.  In this case degeneracy occurs when both elements lie on a chain where the two endpoints forward project onto the same element on the quantifying chain and backward project onto the same element on the quantifying chain.  The interval is called \textit{purely chain-like} if both elements of its quantification pair are positive and equal.

An interval is called \textit{antichain-like} iff the elements of its quantifying pair are of opposite sign, with $(0,0)$ as a degenerate case where the endpoint elements are incomparable, but forward project onto the same element on the quantifying chain and backward project onto the same element on the quantifying chain.  If both elements of the quantifying pair have the same value and opposite signs, then the interval is called \textit{purely antichain-like}.

 An interval is called \textit{projection-like} iff only one of the elements of its quantifying pair is zero, with $(0,0)$ as a degenerate case where the interval itself is degenerate such that the endpoints of the interval are identical and lie on the quantifying chain.  

We will use this quantification technique for two chains in a poset of events instead of one and see how this constraint gives rise to some geometrical features such as dimensionality, directionality and subspaces.

\section{Basic Geometrical Structures}
In this section we will consider quantification with two embedded observers.  First we need to study the relation between an event and a pair of observers via chain projection.  We will see how the order in projections gives rise to collinearity, directionality and subspaces.  

\subsection{Collinearity and Subspaces}
The order in chain projection induces structure in the partially-ordered set.  A pair of distinct chains $\textbf{P}, \textbf{Q}\in \Pi$ and an event $x\in \Pi$ can be \textit{forward collinear} iff the forward projections of event $x$ onto $\textbf{Q}$ can be found by first forward projecting $x$ onto $\textbf{P}$ and then either forward or backward projecting onto $\textbf{Q}$, or \textit{backward collinear} iff the backward projections of event $x$ onto $\textbf{Q}$ can be found by first forward projecting $x$ onto $\textbf{P}$ and then either forward or backward projecting onto $\textbf{Q}$, or \textit{collinear} iff event $x$ is both forward and backward collinear with its projections onto the two chains \cite{KB1}\cite{KB2}.

\begin{figure}[t]
\centering
\includegraphics[width=0.90
\textwidth]{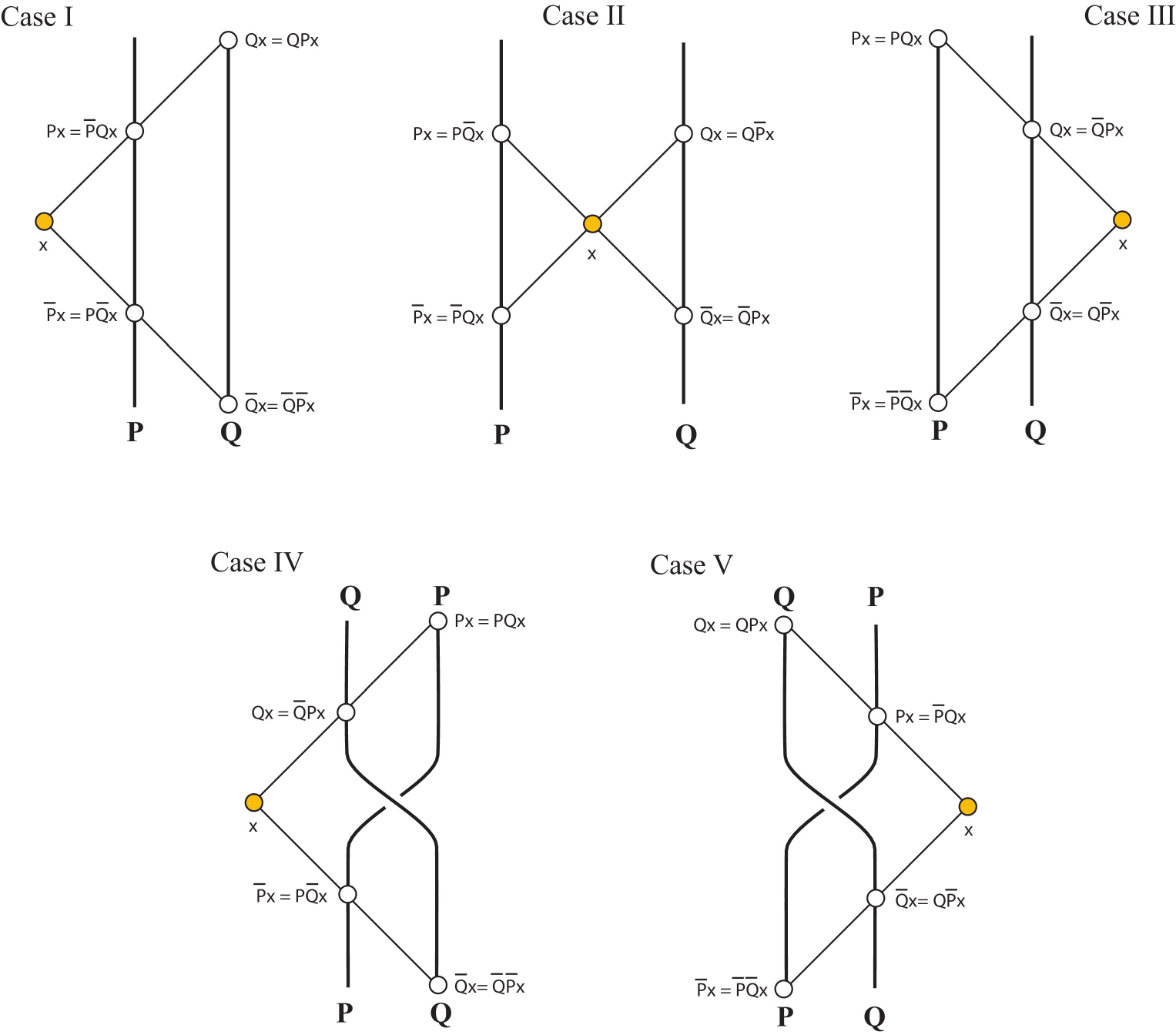}  
\caption{All possible ways in which an event $x$ is collinear with its projections onto two distinct finite chains $\textbf{P}$ and $\textbf{Q}$.  Cases I-III illustrate the concept of proper collinearity which enables us to define directionality and betweenness.}
\label{fig:Figure2}
\end{figure}

One can show \footnote{The proof is given in the appendix.} that the following five cases are the only possible cases in which an event $x$ and two distinct chains $\textbf{P}$ and $\textbf{Q}$ in a partially-ordered set can be collinear as shown in Figs. \ref{fig:Figure2}

Case I:
\begin{equation*}
Px=\overline{P}Qx  \ \ \ \ \ \ \  Qx= QPx 
\end{equation*}
\begin{equation*}
\overline{P}x=P\overline{Q}x  \ \ \ \ \ \ \  \overline{Q}x=\overline{Q}\overline{P}x
\end{equation*}

Case II:
\begin{equation*}
Px=P\overline{Q}x  \ \ \ \ \ \ \  Qx= Q\overline{P}x 
\end{equation*}
\begin{equation*}
\overline{P}x=\overline{P}Qx  \ \ \ \ \ \ \  \overline{Q}x=\overline{Q}Px
\end{equation*}

Case III:
\begin{equation*}
Px=PQx  \ \ \ \ \ \ \  Qx= \overline{Q}Px 
\end{equation*}
\begin{equation*}
\overline{P}x=\overline{P}\overline{Q}x  \ \ \ \ \ \ \  \overline{Q}x=Q\overline{P}x
\end{equation*}

Case IV:
\begin{equation*}
Px=PQx  \ \ \ \ \ \ \  Qx= \overline{Q}Px 
\end{equation*}
\begin{equation*}
\overline{P}x=P\overline{Q}x  \ \ \ \ \ \ \  \overline{Q}x=\overline{Q}\overline{P}x
\end{equation*}

Case V:
\begin{equation*}
Px=\overline{P}Qx  \ \ \ \ \ \ \  Qx= QPx 
\end{equation*} 
\begin{equation*}
\overline{P}x=\overline{P}\overline{Q}x  \ \ \ \ \ \ \  \overline{Q}x=Q\overline{P}x.
\end{equation*}

The invariance of the Cases I, II and III with respect to interchanging forward and backward projections leads us to a more specific definition of collinearity called \textit{proper collinearity} where an event $x$ is said to be properly collinear with its projections onto two distinct finite chains $\textbf{P}$ and $\textbf{Q}$ iff it is collinear with its projections onto the two chains and those projections are invariant with respect to reversing the ordering relation.  The generalization of proper collinearity to three finite chains $\textbf{X}, \textbf{P}, \textbf{Q} \in \Pi$ is made iff each event $x\in \textbf{X}$ is properly collinear with its projections onto $\textbf{P}$ and $\textbf{Q}$, and these projections constitute a surjective map from $\textbf{X}$ onto the finite subchains defined by the closed interval $[Px_{max},\overline{P}x_{min}]$ quantified by $\textbf{P}$, and interval $[Qx_{max},\overline{Q}x_{min}]$ quantified by chain $\textbf{Q}$, denoted by $[Px_{max},\overline{P}x_{min}]_{\textbf{P}}$ and $[Qx_{max},\overline{Q}x_{min}]_{\textbf{Q}}$.

\begin{figure}[t]
\centering
\includegraphics[width=0.50%
\textwidth]{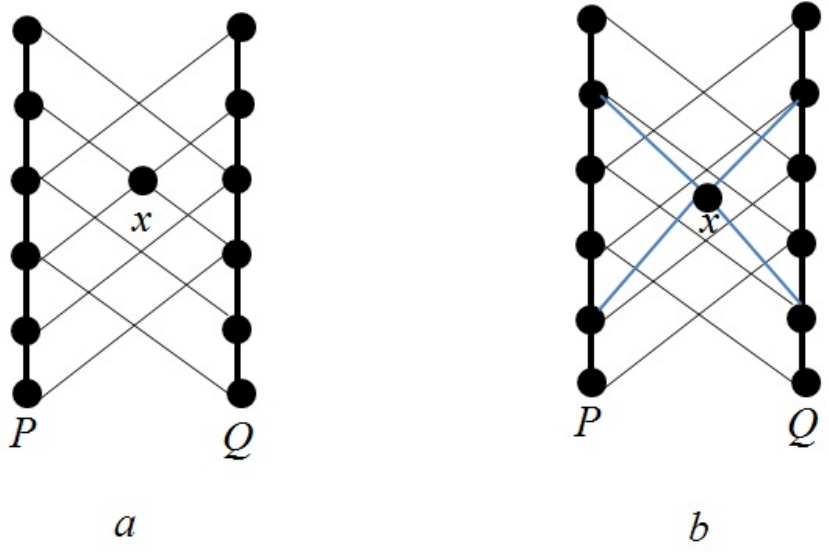}  
\caption{Potential relationships between event $x$ and a subspace $\left\langle \textbf{PQ} \right\rangle$ formed by two chains $\textbf{P}$ and $\textbf{Q}$.  a) Event $x$ is in the subspace formed by $\textbf{P}$ and $\textbf{Q}$, denoted $x\in \left\langle \textbf{PQ} \right\rangle$.  b) Event $x$ is not in the subspace formed by $\textbf{P}$ and $\textbf{Q}$, denoted $x\notin \left\langle \textbf{PQ} \right\rangle$.}
\label{fig:Figure3}
\end{figure}

Collinearity divides the partially-ordered set into two equivalence classes; events that are collinear with two distinct finite chains, and events that are not collinear with two distinct finite chains.  This results in the definition of a \textit{subspace}.  Events that are properly collinear with the two distinct finite chains $\textbf{P}$ and $\textbf{Q}$ reside in a \textit{discrete subspace} defined by the two chains, denoted by $\left\langle \textbf{PQ}\right\rangle$ shown in Fig. \ref{fig:Figure3} (a).  Chains $\textbf{P}$ and $\textbf{Q}$ are defined to be elements of the subspace \cite{KB1}\cite{KB2}.

Collinearity of an event $x$ and two distinct observers can be investigated further by looking at the order in the projections and see how this induces directionality.

\subsection{Directionality}

In addition to discrete subspaces, collinearity gives rise to the concept of \textit{directionality} and \textit{betweenness} in a partially-ordered set.
Event $x$ is said to be on the $\textbf{P}$ side of two distinct finite chains $\textbf{P}$ and $\textbf{Q}$ if the forward and backward projections of event $x$ onto chain $\textbf{Q}$ are first found on chain $\textbf{P}$ denoted $\textbf{Q}x=\textbf{Q}\textbf{P}x$ and $\overline{\textbf{Q}}x=\overline{\textbf{Q}}\overline{\textbf{P}}x$ for the forward and backward projections respectively, denoted $x|\textbf{P}|\textbf{Q}$ shown in Fig. \ref{fig:Figure2} (Case I).  Similarly if projections of event $x$ onto chain $\textbf{P}$ is first found on chain $\textbf{Q}$ denoted $\textbf{P}x=\textbf{PQ}x$ and $\overline{\textbf{P}}x=\overline{\textbf{P}}\overline{\textbf{Q}}x$ for the forward and backward projections respectively, then $x$ is said to be on the \textbf{$\textbf{Q}$-side} of the two finite chains, denoted $x|\textbf{Q}|\textbf{P}$ shown in Fig. \ref{fig:Figure2} (Case II) \cite{KB1}\cite{KB2}.  

Event $x$ is said to be \textit{between} $\textbf{P}$ and $\textbf{Q}$ denoted, $\textbf{P}|x|\textbf{Q}$, if event $x$ is first backward projected onto chain $\textbf{P}$ and then forward projected onto chain $\textbf{Q}$, denoted $\textbf{Q}\overline{\textbf{P}}x$, and similarly it is forward projected onto chain $\textbf{P}$ and then backward projected onto chain $\textbf{Q}$, denoted $\overline{\textbf{Q}}\textbf{P}x$ shown in Fig. \ref{fig:Figure2} (Case II).

These notions can be extended to three chains $\textbf{P}$, $\textbf{Q}$ and $\textbf{X}$ when all the events on chain $\textbf{X}$ are properly collinear with their projections on $\textbf{P}$ and $\textbf{Q}$.  
\begin{equation}
\label{eq:1}
\textbf{X}|\textbf{P}|\textbf{Q}\ \ \ \ \textbf{P}|\textbf{X}|\textbf{Q}\ \ \ \ \textbf{X}|\textbf{Q}|\textbf{P}.
\end{equation}
This results in an ordering relation among chains such that 
\begin{equation}
\label{eq:2}
					\textbf{X}|\textbf{P}|\textbf{Q} \Rightarrow 
					\begin{cases}
        \textbf{X}<\textbf{P}<\textbf{Q} \\
        \textbf{Q}<\textbf{P}<\textbf{X}
    \end{cases}  
					 \end{equation}
where $<$ indicates that $\textbf{X}\leq \textbf{P}$ but $\textbf{X}\neq\textbf{P}$.  Thus chains can be ordered, where the direction of ordering is arbitrary.

The subspace $\left\langle \textbf{PQ} \right\rangle$ defined by two finite chains $\textbf{P}, \textbf{Q}\in \Pi$ is two-dimensional with one dimension resulting from the natural causal ordering among sequences of events along chains, and the other resulting from an induced order that is caused by collinearity \cite{KB1}.  To differentiate between the natural ordering and the induced ordering, we say that the subspace $\left\langle \textbf{PQ} \right\rangle$ is 1+1 dimensional \cite{KB1}\cite{KB2}.

A pair of collinear chains that form a subspace can be used as a pair of quantifying embedded observers.  The results of this quantification lead to emergence of spacetime and the Minkowski metric as discussed in detail in our previous work \cite{KB1}.  Here we will review the coordination condition and study different cases as we increase the number of coordinated chains.

\subsection{Coordinated Observers}
Two chains $\textbf{P}$ and $\textbf{Q}$ are said to be \textit{coordinated} over the intervals given by $[\bar{p}_{min},\bar{p}_{max}]_{\textbf{P}}$, $[p_{min},p_{max}]_{\textbf{P}}$ and $[\bar{q}_{min},\bar{q}_{max}]_{\textbf{Q}}$, $[q_{min},q_{max}]_{\textbf{Q}}$ iff $[\bar{p}_{min},\bar{p}_{max}]_{\textbf{P}}$ forward projects onto $[q_{min},q_{max}]_{\textbf{Q}}$ and $[p_{min},p_{max}]_{\textbf{P}}$ backward projects onto $[\bar{q}_{min},\bar{q}_{max}]_{\textbf{Q}}$ such that the projections have one-to-one correspondence, and if the length of a closed interval on $\textbf{P}$ is equal to the length of its image on $\textbf{Q}$ and vice versa.  Moreover, we postulate that if two chains agree on the quantification of each other's intervals, then they must agree on the quantification of all the intervals they both can quantify \cite{KB1}\cite{KB2}.

The set of coordinated chains $\textbf{P}$ and $\textbf{Q}$  form a 1+1 dimensional subspace denoted by $\left\langle \textbf{PQ}\right\rangle$ which includes all elements collinear with these two chains as shown in Fig. \ref{fig:Figure3} (a).  Thus the two coordinated chains can quantify all events that belong to this subspace \cite{KB1}\cite{KB2}.  

It can be shown \cite{KB1} that any generalized interval $[x,y]$ that belongs to the 1+1 dimensional subspace $\left\langle \textbf{PT}\right\rangle$ defined by a set of coordinated chains $\textbf{P}$, $\textbf{Q}$ and $\textbf{T}$ can be quantified consistently with one of the following pairs:
  
\begin{align*}
I.\ \ \ [x,y]_{\textbf{P}} & = (v_{\textbf{P}}(Py)-v_{\textbf{P}}(Px),v_{\textbf{P}}(\overline{P}y)-v_{\textbf{P}}(\overline{P}x))_{\textbf{P}} \\
 & = (p_{y}-p_{x},\overline{p}_{y}-\overline{p}_{x})=(\Delta p,\Delta \overline{p}),
\end{align*}
when the interval is on one side of the chain $\textbf{P}$;
\begin{align*}
II.\ \ \  [x,y]_{\textbf{P}} & = (v_{\textbf{P}}(Py)-v_{\textbf{P}}(\overline{P}x),v_{\textbf{P}}(\overline{P}y)-v_{\textbf{P}}(Px))_{\textbf{P}} \\
 & = (p_{y}-\overline{p}_{x},\overline{p}_{y}-p_{x}),
\end{align*}
when the interval is on both sides of the chain $\textbf{T}$; and
\begin{equation*}
III.\ \ \ [x,y]_{\textbf{PQ}}=(p_{y}-p_{x},q_{y}-q_{x})=(\Delta p, \Delta q)_{\textbf{PQ}},
\end{equation*}
when both events of an interval are situated between the two quantifying chains, $\textbf{P}|x|\textbf{Q}$ and $\textbf{P}|y|\textbf{Q}$.

Next, we will consider a special case of two pairs of coordinated chains and use that special case to derive a discrete version of the Pythagorean theorem. 

\subsection{Orthogonal Subspaces}
Here we will discuss a special case of the two subspaces that will motivate a more general concept of orthogonal subspaces.

\begin{figure}[t]
\centering
\includegraphics[width=0.35
\textwidth]{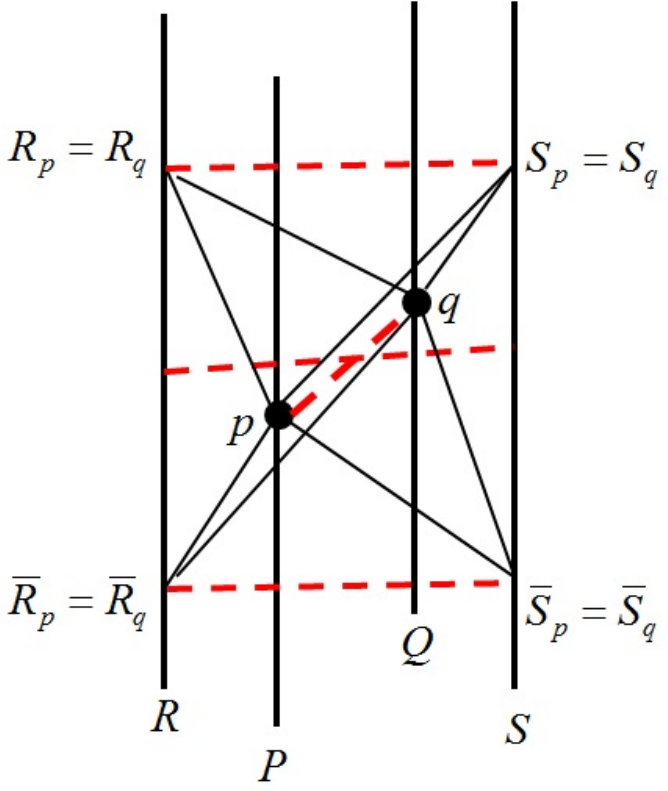}  
\caption{A motivating example for the concept of orthogonality.  The chains $\textbf{P}$ and $\textbf{Q}$ are coordinated and the chains $\textbf{R}$ and $\textbf{S}$ are also coordinated.  The pair $(0,0)$ is the quantification pair for interval $[p,q]$ as quantified by the chains $\textbf{R}$ and $\textbf{S}$.  Thus the two subspaces $\left\langle \textbf{PQ}\right\rangle$ and $\left\langle \textbf{RS}\right\rangle$ are said to be orthogonal.}
\label{fig:Figure4}
\end{figure}

Consider two pairs of coordinated chains $\textbf{PQ}$ and $\textbf{RS}$ and their corresponding subspaces $\left\langle \textbf{PQ}\right\rangle$ and $\left\langle \textbf{RS}\right\rangle$, respectively.  Consider events $p\in \textbf{P}$ and $q\in \textbf{Q}$ such that $[p,q]$ is a purely antichain-like interval as quantified by $\textbf{PQ}$, so that the quantification pair for interval $[p,q]$ with respect to $\overline{\textbf{PQ}}$ is given by 
\begin{equation}
\label{eq:3}
[p,q]_{\textbf{PQ}}=(v_{\textbf{P}}(Pq)-v_{\textbf{P}}(p),v_{\textbf{Q}}(q)-v_{\textbf{Q}}(Qp))=(\Delta,-\Delta).
\end{equation}
Now consider a special case where the two coordinated chains $\textbf{RS}$ quantify events $p$ and $q$ with the same valuations so that, $Rp=Rq$, $Sp=Sq$, $\overline{R}p=\overline{R}q$ and $\overline{S}p=\overline{S}q$.  This results in the following antichain-like degenerate quantification pair for the interval $[p,q]$ with respect to chains $\textbf{RS}$
\begin{equation}
\label{eq:4}
[p,q]_{\textbf{RS}}=(v_{\textbf{P}}(Rq)-v_{\textbf{P}}(Rp),v_{\textbf{S}}(Sq)-v_{\textbf{S}}(Sp))=(0,0).
\end{equation}
Note that $\textbf{PQ}$ also quantifies both $[\overline{R}p,\overline{S}p]$ and $[Rp,Sp]$ in an antichain-like degenerate fashion,
\begin{equation}
\label{eq:5}
[\overline{R}p,\overline{S}p]_{\textbf{PQ}}=(v_{\textbf{P}}(\overline{S}p)-v_{\textbf{P}}(\overline{R}p),v_{\textbf{Q}}(\overline{S}p)-v_{\textbf{Q}}(\overline{R}p))=(0,0),
\end{equation}
\begin{equation}
\label{eq:6}
[Rp,Sp]_{\textbf{PQ}}=(v_{\textbf{P}}(Sp)-v_{\textbf{P}}(Rp),v_{\textbf{Q}}(Sp)-v_{\textbf{Q}}(Rp))=(0,0),
\end{equation}
where $v_{\textbf{P}}$ is the valuation assigned by chain $\textbf{P}$.
In this situation the two subspaces $\overline{\textbf{PQ}}$ and $\overline{\textbf{RS}}$ are said to be orthogonal to one another as shown in Fig. \ref{fig:Figure4}.
Note that this is not a definition of orthogonality.  This represents a sufficient condition but not a necessary one.  A definition of orthogonality will be given in section 7.

We will use two pairs of orthogonal coordinated chains that were introduced here to derive the Pythagorean theorem.

\subsection{Pythagorean Theorem}
\begin{figure}[t]
\centering
\includegraphics[width=0.40
\textwidth]{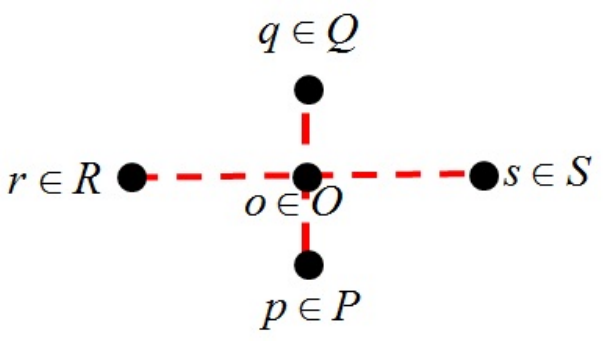}  
\caption{A geometric view of the two orthogonal subspaces $\left\langle \textbf{PQ}\right\rangle$ and $\left\langle \textbf{RS}\right\rangle$ where the dots show the geometric view of the chains and the dotted lines show the antichain-like intervals with their endpoints on the chains.  The addition of the interval scalars for orthogonal subspaces gives the Pythagorean theorem in the poset picture.}
\label{fig:Figure5}
\end{figure}

We use a special configuration of chains to prove the Pythagorean theorem in the poset picture.  Consider two sets of coordinated chains $\textbf{P}|\textbf{O}|\textbf{Q}$ and $\textbf{R}|\textbf{O}|\textbf{S}$ such that they form two orthogonal subspaces and three events $p\in \textbf{P}$, $r\in \textbf{R}$ and $o\in \textbf{O}$ as shown in Fig. \ref{fig:Figure5}.  The corresponding antichain-like intervals $[p,o]$, $[o,r]$ and $[p,r]$ will then be such that
\begin{equation}
\label{eq:7}
[p,o]\uplus [o,r]=[p,r],
\end{equation}
where $\uplus$ is the operation that maps the two intervals onto the single interval.  For the quantifying pairs, we write
\begin{equation}
\label{eq:8}
(p_{o}-p,o-o_{p})_{\textbf{PO}}\oplus (o_{r}-o,r-r_{o})_{\textbf{OR}} \sim (p_{r}-p,r-r_{p})_{\textbf{PR}},
\end{equation}
where $p_{o}-p=-(o-o_{p})$, $o_{r}-o=-(r-r_{o})$, $p_{r}-p=(r-r_{p})$, $\oplus$ is a function that combines the orthogonal pairs and $\sim$ is a relation that is the result of combining the quantification pairs of the orthogonal intervals since combining the pairs $(p_{o}-p,o-o_{p})_{\textbf{PO}}$ and $(o_{r}-o,r-r_{o})_{\textbf{OR}}$ in this way is not equal to $(p_{r}-p,r-r_{p})_{\textbf{PR}}$.  Let $p_{o}-p=\Delta a$, $o_{r}-o=\Delta b$ and $p_{r}-p=\Delta c$. Then Eq. (\ref{eq:8}) can be rewritten as
\begin{equation}
\label{eq:9}
(\Delta a,-\Delta a)\oplus (\Delta b,-\Delta b) \sim (\Delta c,-\Delta c).
\end{equation}

Previously we have shown that any quantifiable interval can be quantified by a pair as well as a scalar \cite{KB1}.  We have shown that the scalar quantification of a generalized interval that is quantified by the pair $(\Delta p,\Delta \overline{p})$ with respect to a chain $P$ is given by $\Delta p \Delta\overline{p}$, which is the product of the two elements of the quantifying pair.  Equivalently, for a generalized interval quantified by the pair $(\Delta p,\Delta q)$ with respect to coordinated chains $\textbf{P}$ and $\textbf{Q}$, the scalar quantification is given by $\Delta p\Delta q$.  This also resulted in a relation among orthogonal intervals such that for orthogonal intervals $[p,o]$ and $[o,r]$, using this result, we have 
\begin{equation}
\label{eq:10}
(\Delta a)^{2}+(\Delta b)^{2}=(\Delta c)^{2}
\end{equation}
among their corresponding interval scalars, which is the familiar Pythagorean theorem applied to antichain-like intervals.
Increasing the number of embedded observers from two to three in the special case where the embedded observers belong to orthogonal subspaces results in the Pythagorean theorem.  In the next section we will study two other special cases of adding more quantifying chains and discuss the results.

\section{Simplices}
We can now show how geometrical shapes, more specifically \textit{simplices}, arise in a partially-ordered set of events when we have more than a pair of quantifying embedded observers.  We will consider the cases where we have three and then four observers and discuss each result and the generalization to N observers.  

\subsection{Discrete Equilateral Triangle}
We begin by considering a set of three coordinated chains in two different configurations; when all are collinear and, when they are collinear pairwise.  Then we discuss how each configuration is related to a geometric shape.
 
\begin{figure}[t]
\centering
\includegraphics[width=0.70
\textwidth]{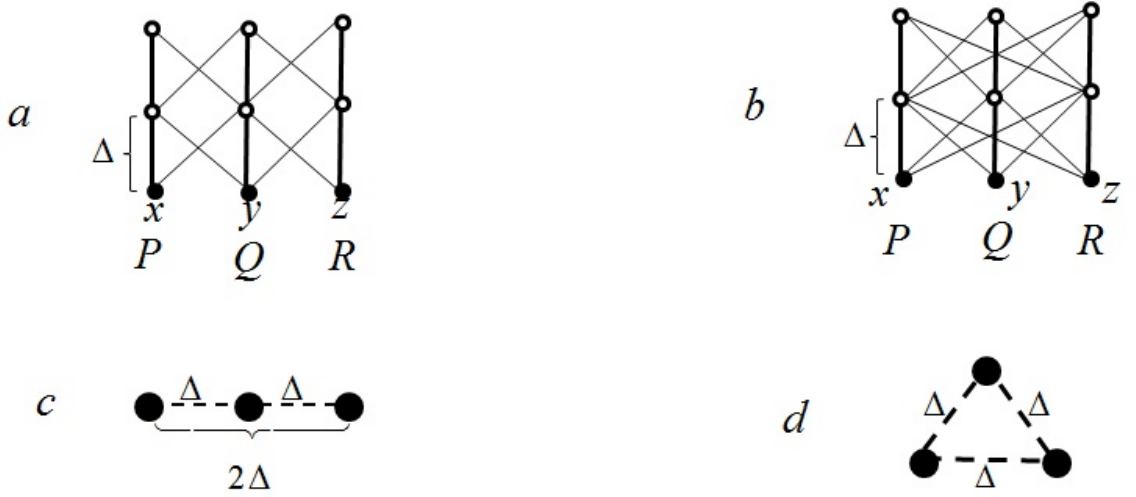}  
\caption{Two cases of three coordinated chains $\textbf{P}$, $\textbf{Q}$ and $\textbf{R}$.  In Case I shown in (a) they are all collinear and in Case II shown in (b) they are collinear pairwise.  The intervals $[x_{i},y_{i}]$, $[y_{i},z_{i}]$ and $[x_{i},z_{i}]$ are quantified with respect to pairs of chains $\textbf{PQ}$, $\textbf{QR}$ and $\textbf{PR}$ respectively in both cases.  This shows that in Case II intervals form a discrete equilateral triangle shown in (d).  The dots in Figs. (c), (d) show the geometric view of the chains and their corresponding events and the dotted lines show the antichain-like intervals formed by the events.}
\label{fig:Figure6}
\end{figure}

Consider three coordinated chains $\textbf{P}$, $\textbf{Q}$ and $\textbf{R}$ with events $x_{i}\in \textbf{P}$, $y_{i}\in \textbf{Q}$ and $z_{i}\in \textbf{R}$.  Two possible configurations for this set of three coordinated chains are: 

\textit{Case I}: Coordinated chains $\textbf{P}$, $\textbf{Q}$ and $\textbf{R}$ are collinear as shown in Fig. \ref{fig:Figure6} (a), \textit{Case II}: Coordinated chains $\textbf{P}$, $\textbf{Q}$ and $\textbf{R}$ are collinear pairwise as shown in Fig. \ref{fig:Figure6} (b).  
According to the configuration in Case \textit{I}, the finite chain $\textbf{Q}$ is between the finite chains $\textbf{P}$ and $\textbf{R}$, $\textbf{P}|\textbf{Q}|\textbf{R}$.  Consider the situation where $\forall i$, $[x_{i},y_{i}]_{\textbf{PQ}}$, $[y_{i},z_{i}]_{\textbf{QR}}$ and $[x_{i},z_{i}]_{\textbf{PR}}$ are all pure antichain-like intervals.

\textit{Case I}:
Quantifying intervals $[x_{i},y_{i}]$, $[y_{i},z_{i}]$ and $[x_{i},z_{i}]$ in Case \textit{I} with respect to chains $\textbf{PQ}$, $\textbf{QR}$ and $\textbf{PR}$ respectively, gives 
\begin{equation}
\label{eq:11} 
[x_{i},y_{i}]_{\textbf{PQ}}=(p_{y_{i}}-p_{x_{i}},q_{y_{i}}-q_{x_{i}})_{\textbf{PQ}}=(\Delta,-\Delta). 
\end{equation}

In general, the quantification pair of an interval can be decomposed using a relation called the \textit{symmetric-antisymmetric decomposition}, where an interval pair $(\Delta p,\Delta q)_{\textbf{PQ}}$ can be rewritten as a component-wise sum of a symmetric pair and an antisymmetric pair such that \cite{KB1}\cite{KB2} 
\begin{equation}
\label{eq:12}
(\Delta p, \Delta q)_{\textbf{PQ}}=\big(\frac{\Delta p+\Delta q}{2},\frac{\Delta p+\Delta q}{2}\big)_{\textbf{PQ}}+\big(\frac{\Delta p-\Delta q}{2},\frac{\Delta q-\Delta p}{2}\big)_{\textbf{PQ}}.
\end{equation}
Introduce the change of variables based on the length as measured along chains and the distance as measured between coordinated chains
\begin{equation}
\label{eq:13}
\Delta t = \frac{\Delta p+\Delta q}{2},
\end{equation}
\begin{equation}
\label{eq:14}
\Delta x = \frac{\Delta p-\Delta q}{2}.
\end{equation}
Rewriting Eq. (\ref{eq:12}) in terms of the new variables gives
\begin{equation}
\label{eq:15}
(\Delta p,\Delta q)=(\Delta t,\Delta t)+(\Delta x,-\Delta x).
\end{equation}

We now use this format and rewrite Eq. (\ref{eq:11}) in terms of the symmetric-antisymmetric pairs $\Delta t$ and $\Delta x$
\begin{equation}
\label{eq:16}
[x_{i},y_{i}]_{\textbf{PQ}}=\big(\frac{\Delta -\Delta}{2},\frac{\Delta -\Delta}{2}\big)+\big(\frac{\Delta +\Delta}{2},\frac{-\Delta -\Delta}{2}\big)=(0,0)+(\Delta,-\Delta).
\end{equation}  
Coordination condition demands that event $i$ on chain $\textbf{P}$ projects onto event $i+1$ on chain $\textbf{Q}$.  This gives $\Delta =1$.  Similarly, for interval $[y_{i},z_{i}]$ we have
\begin{equation}
\label{eq:17}
[y_{i},z_{i}]_{\textbf{QR}}=(q_{z_{i}}-q_{y_{i}},r_{z_{i}}-r_{y_{i}})_{\textbf{QR}}=(1,-1),
\end{equation}
and, for interval $[x_{i},z_{i}]$ we have
\begin{equation}
\label{eq:18}
[x_{i},z_{i}]_{\textbf{PR}}=(p_{z_{i}}-p_{x_{i}},r_{z_{i}}-r_{x_{i}})_{\textbf{PR}}=(2,-2). 
\end{equation}
Geometric view of this case is shown in Fig. \ref{fig:Figure6} (c), where the dots refer to the events $x_{i}$, $y_{i}$ and $z_{i}$ and the dotted lines show the antichain-like intervals $[x_{i},y_{i}]$, $[y_{i},z_{i}]$ and $[x_{i},z_{i}]$.  Therefore collinearity imposed on the three coordinated chains makes the intervals $[x_{i},y_{i}]$, $[y_{i},z_{i}]$ and $[x_{i},z_{i}]$ aligned such that
\begin{equation}
\label{eq:19}
[x_{i},y_{i}]_{\textbf{PQ}}=[y_{i},z_{i}]_{\textbf{QR}}=\frac{1}{2}[x_{i},z_{i}]_{\textbf{PR}}.
\end{equation} 
It is important to mention here that although the quantifications are with respect to different pairs of chains, since the chains are all coordinated they can compare quantifications.

\textit{Case II}:
In this case the coordinated chains are collinear pairwise.  The quantification pairs of the intervals $[x_{i},y_{i}]$, $[y_{i},z_{i}]$ and $[x_{i},z_{i}]$ with respect to chains $\textbf{PQ}$ and $\textbf{QR}$ and $\textbf{PR}$ would be
\begin{equation} 
\label{eq:20}
[x_{i},y_{i}]_{\textbf{PQ}}=(p_{y_{i}}-p_{x_{i}},q_{y_{i}}-q_{x_{i}})_{\textbf{PQ}}=(1,-1), 
\end{equation}
or in terms of $\Delta t$ and $\Delta x$ we have
\begin{equation}
\label{eq:21}
[x_{i},y_{i}]_{\textbf{PQ}}=\big(\frac{\Delta -\Delta}{2},\frac{\Delta -\Delta}{2}\big)+\big(\frac{\Delta +\Delta}{2},\frac{-\Delta -\Delta}{2}\big)=(0,0)+(1,-1).
\end{equation}  
For intervals $[y_{i},z_{i}]$ and $[x_{i},z_{i}]$, we have
\begin{equation}
\label{eq:22}
[y_{i},z_{i}]_{\textbf{QR}}=(q_{z_{i}}-q_{y_{i}},r_{z_{i}}-r_{y_{i}})_{\textbf{QR}}=(1,-1)
\end{equation}      
\begin{equation}
\label{eq:23}
[x_{i},z_{i}]_{\textbf{PR}}=(p_{z_{i}}-p_{x_{i}},r_{z_{i}}-r_{x_{i}})_{\textbf{PR}}=(1,-1). 
\end{equation}
The geometric view of Case \textit{II} is shown in Fig. \ref{fig:Figure6} (d), where each chain is equidistant from every other chain.  This gives the relation among the intervals as 
\begin{equation}
\label{eq:24}
[x_{i},y_{i}]_{\textbf{PQ}}=[y_{i},z_{i}]_{\textbf{QR}}=[x_{i},z_{i}]_{\textbf{PR}},
\end{equation}
which shows that the intervals $[x_{i},y_{i}]$, $[y_{i},z_{i}]$ and $[x_{i},z_{i}]$ form a \textit{discrete equilateral triangle}.

This demonstrates that chain projection can be used to order chains in such a way that they can result in the formation of geometric shapes.

We will increase the number of quantifying chains from three to four and discuss the results next.

\subsection{Discrete Tetrahedron}
\begin{figure}[t]
\centering
\includegraphics[width=0.70
\textwidth]{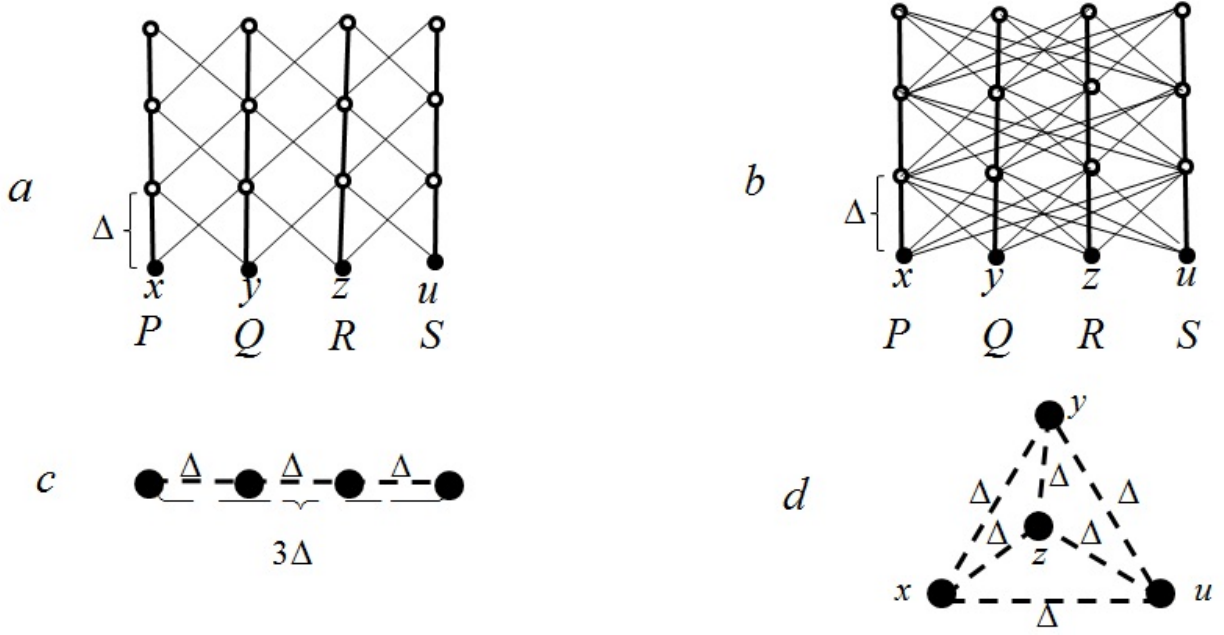}  
\caption{Four coordinated chains $\textbf{P}$, $\textbf{Q}$, $\textbf{R}$ and $\textbf{S}$ in two cases; Case \textit{I}: all four coordinated chains are collinear, and Case \textit{II}: the four coordinated chains are collinear pairwise.  Quantifying intervals $[x_{i}, y_{i}]$, $[y_{i}, z_{i}]$, $[z_{i}, u_{i}]$, $[x_{i}, z_{i}]$, $[y_{i}, u_{i}]$ and $[x_{i}, u_{i}]$ with respect to pairs of chains $\textbf{PQ}$, $\textbf{QR}$, $\textbf{RS}$, $\textbf{PR}$, $\textbf{QS}$ and $\textbf{PS}$ respectively, results in purely antichain-like intervals that are aligned in Case \textit{I} as shown in (c), whereas in Case \textit{II} they form a discrete tetrahedron shown in (d).}
\label{fig:Figure7}
\end{figure}

Now let us consider four coordinated chains in both cases discussed above where \textit{Case I} refers to the set of coordinated chains that are all collinear, and \textit{Case II} refers to the set of four coordinated chains that are collinear pairwise.  Consider four events $x_{i}\in \textbf{P}$, $y_{i}\in \textbf{Q}$, $z_{i}\in \textbf{R}$ and $u_{i}\in \textbf{S}$, such that, $\forall i$, they form antichain-like intervals $[x_{i}, y_{i}]$, $[y_{i}, z_{i}]$, $[z_{i}, u_{i}]$, $[x_{i}, z_{i}]$, $[y_{i}, u_{i}]$ and $[x_{i}, u_{i}]$ as shown in Figs. \ref{fig:Figure7} (a), \ref{fig:Figure7} (b).

We will quantify purely antichain-like intervals $[x_{i},y_{i}]_{\textbf{PQ}}$, $[y_{i},z_{i}]_{\textbf{QR}}$, $[x_{i},z_{i}]_{\textbf{PR}}$, $[z_{i},u_{i}]_{\textbf{RS}}$, $[y_{i},u_{i}]_{\textbf{QS}}$ and $[x_{i},u_{i}]_{\textbf{PS}}$ with respect to the pair of coordinated chains shown in their subscripts for each case.

\textit{Case I}:    

\begin{equation}
\label{eq:25}
[x_{i},y_{i}]_{\textbf{PQ}}=(p_{y_{i}}-p_{x_{i}},q_{y_{i}}-q_{x_{i}})=(\Delta,-\Delta),
\end{equation}  
where $\Delta =1$ due to the coordination condition.
Similarly for the interval $[y_{i},z_{i}]$ quantified by the coordinated chains $\textbf{QR}$ we have
\begin{equation}
\label{eq:26}
[y_{i},z_{i}]_{\textbf{QR}}=(q_{z_{i}}-q_{y_{i}},r_{z_{i}}-r_{y_{i}})=(1,-1),
\end{equation}
and for the interval $[z_{i},u_{i}]$ the quantification pair with respect to the coordinated chains $\textbf{RS}$ we have
\begin{equation}
\label{eq:27}
[z_{i},u_{i}]_{\textbf{RS}}=(r_{u_{i}}-r_{z_{i}},s_{u_{i}}-s_{z_{i}})=(1,-1).
\end{equation}
The quantification pair for interval $[x_{i},z_{i}]$ would be
\begin{equation}
\label{eq:28}
[x_{i},z_{i}]_{\textbf{PR}}=(p_{z_{i}}-p_{x_{i}},r_{z_{i}}-r_{x_{i}})=(2,-2).
\end{equation}
Interval $[y_{i},u_{i}]$ with respect to the coordinated chains $\textbf{QS}$ can be quantified as
\begin{equation}
\label{eq:29}
[y_{i},u_{i}]_{\textbf{QS}}=(q_{u_{i}}-q_{y_{i}},s_{u_{i}}-s_{y_{i}})=(2,-2).
\end{equation}
Next consider interval $[x_{i},u_{i}]$ with respect to the coordinated chains $\textbf{PS}$.  In Case \textit{I}  the quantification pair is given by 
\begin{equation}
\label{eq:30}
[x_{i},u_{i}]_{\textbf{PS}}=(p_{u_{i}}-p_{x_{i}},s_{u_{i}}-s_{x_{i}})=(3,-3).
\end{equation}
Since the chains are all coordinated they can compare quantifications.  Comparing the quantification results among the pairs of chains in Case \textit{I} gives
\begin{equation}
\label{eq:31}
[x_{i},y_{i}]_{\textbf{PQ}}=[y_{i},z_{i}]_{\textbf{QR}}=[z_{i},u_{i}]_{\textbf{RS}}=\frac{1}{2}[y_{i},u_{i}]_{\textbf{QS}}=\frac{1}{2}[x_{i},z_{i}]_{\textbf{PR}}=\frac{1}{3}[x_{i},u_{i}]_{\textbf{PS}}.
\end{equation}

\textit{Case II:}

\begin{equation*}
[x_{i},y_{i}]_{\textbf{PQ}}=(p_{y_{i}}-p_{x_{i}},q_{y_{i}}-q_{x_{i}})=(\Delta,-\Delta)=(1,-1)
\end{equation*} 
\begin{equation*}
[y_{i},z_{i}]_{\textbf{QR}}=(q_{z_{i}}-q_{y_{i}},r_{z_{i}}-r_{y_{i}})=(1,-1)
\end{equation*}
\begin{equation*}
[z_{i},u_{i}]_{\textbf{RS}}=(r_{u_{i}}-r_{z_{i}},s_{u_{i}}-s_{z_{i}})=(1,-1)
\end{equation*}
\begin{equation*}
[x_{i},z_{i}]_{\textbf{PR}}=(p_{z_{i}}-p_{x_{i}},r_{z_{i}}-r_{x_{i}})=(1,-1)
\end{equation*}
\begin{equation*}
[y_{i},u_{i}]_{\textbf{QS}}=(q_{u_{i}}-q_{y_{i}},s_{u_{i}}-s_{y_{i}})=(1,-1) 
\end{equation*}
\begin{equation}
\label{eq:32}
[x_{i},u_{i}]_{\textbf{PS}}=(p_{u_{i}}-p_{x_{i}},s_{u_{i}}-s_{x_{i}})=(1,-1).
\end{equation} 
This gives the relation
\begin{equation}
\label{eq:33}
[x_{i},y_{i}]_{\textbf{PQ}}=[y_{i},z_{i}]_{\textbf{QR}}=[z_{i},u_{i}]_{\textbf{RS}}=[y_{i},u_{i}]_{\textbf{QS}}=[x_{i},z_{i}]_{\textbf{PR}}=[x_{i},u_{i}]_{\textbf{PS}}.
\end{equation}
among the quantified intervals.  From this result it can be inferred that the events in Case \textit{II} where the chains are equidistant, are related to a \textit{discrete tetrahedron}.  The rise of geometric simplices implies another important result regarding the dimensionality.  The number of discrete subspaces is 1+1 for a set of 3 and 4 chains when they are coordinated and collinear.  However, the number of discrete subspaces with 4 coordinated chains that are collinear pairwise is 3+1 while the number of discrete subspaces is 2+1 when there are 3 coordinated chains that are collinear pairwise.  In this case, adding a chain corresponds to the addition of one spatial dimension.  This can be used to generalize the results from four coordinated chains to N such that having $N$ equidistant coordinated chains corresponds to an $(N-1)+1$ dimensional subspace.  This means that the quantification technique used to quantify the poset of events and the causal relation among the events do not constraint the dimensionality of space. 

Next, we will consider a set of three or more coordinated and collinear observer chains where all belong to the same subspace, called a \textit{fence}.  We will then show how a discrete version of the Parallel postulate can be proved as we consider different configurations of two fences.

\section{Fence}
\begin{figure}[t]
\centering
\includegraphics[width=0.70
\textwidth]{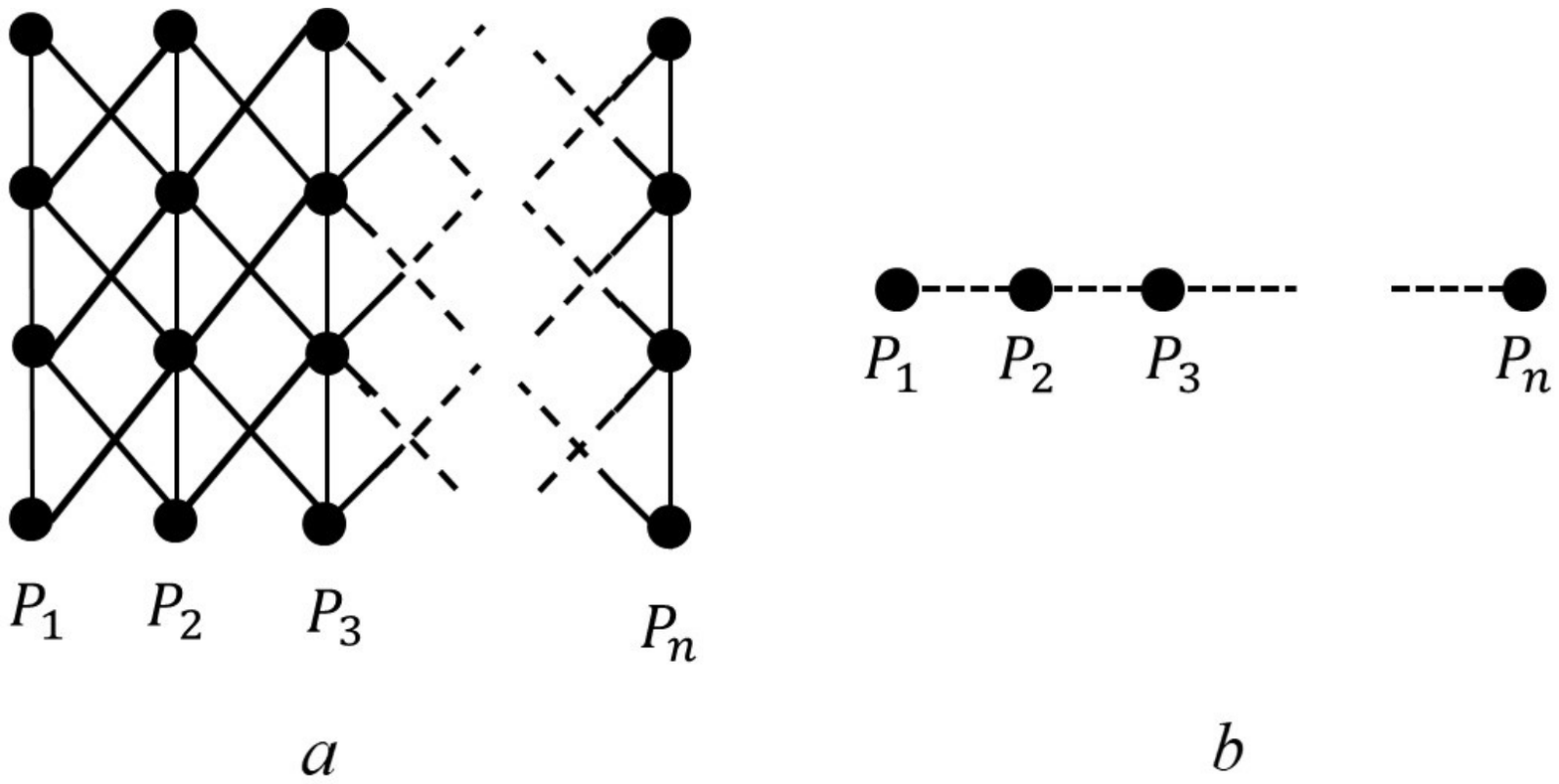}  
\caption{(a) Coordinated and collinear chains $\textbf{P}_{1}, \textbf{P}_{2}, \textbf{P}_{3},..., \textbf{P}_{n}$ form a fence.  (b) Geometric view of fence $\left\|\textbf{P}_{1}\textbf{P}_{n}\right\|$.}
\label{fig:Figure8}
\end{figure}

We define a \textit{fence} as a set of three or more coordinated and collinear chains, denoted by $\left\|\textbf{P}_{1}\textbf{P}_{n}\right\|$ where chains $\textbf{P}_{1}$ and $\textbf{P}_{n}$ are the chains at the two ends of the fence as shown in Fig. \ref{fig:Figure8}.  All chains in the fence $\left\|\textbf{P}_{1}\textbf{P}_{n} \right\|$ belong to the subspace $\left\langle \textbf{P}_{1}\textbf{P}_{n} \right\rangle$.
The distance between the first and the last chains in a fence is denoted by $D(\textbf{P}_{1},\textbf{P}_{n})$.

We will consider different configurations of fences and present a proof of a discrete version of the Parallel postulate.  

\subsection{The Parallel Postulate}
\begin{figure}[t]
\centering
\includegraphics[width=0.70
\textwidth]{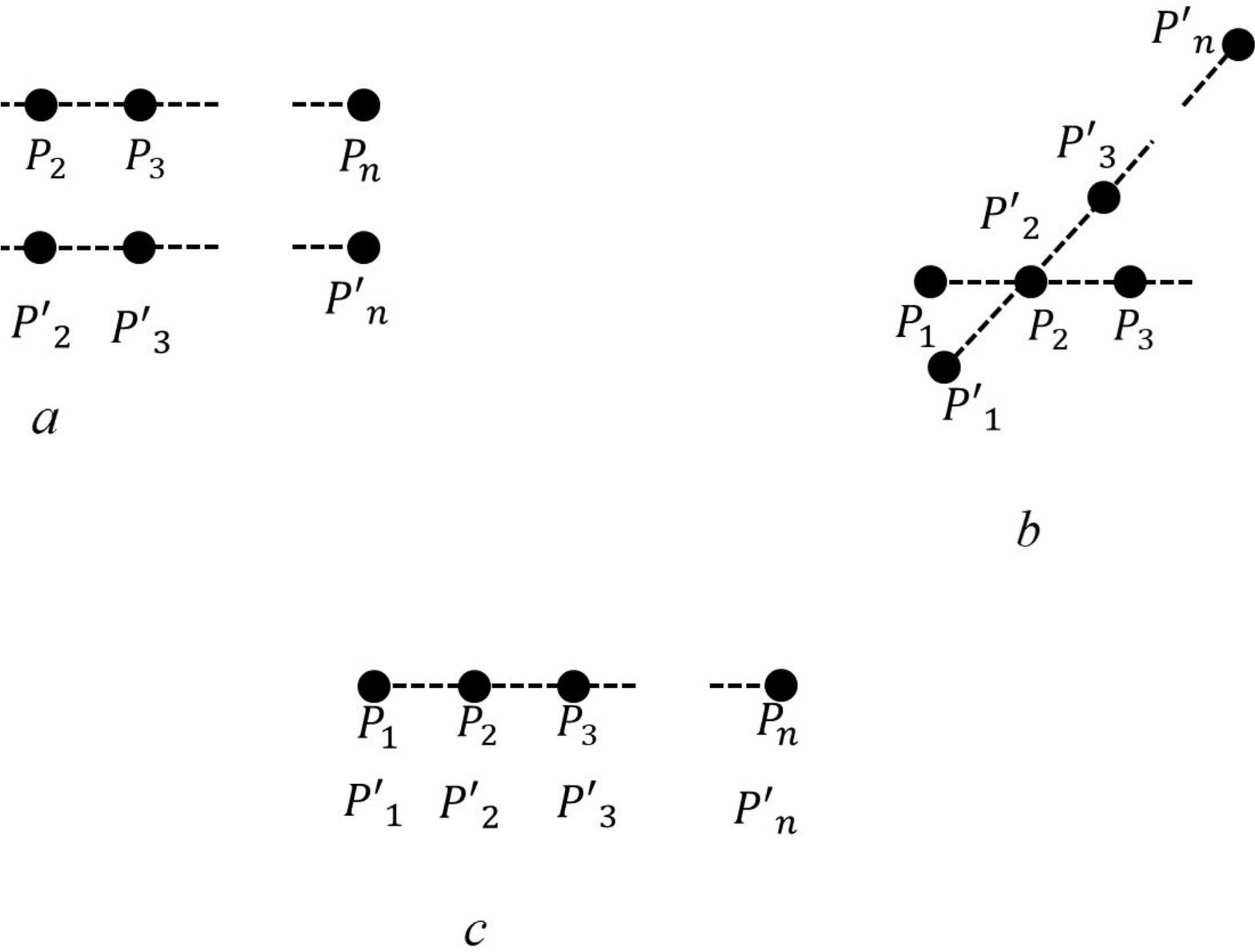}  
\caption{Geometric view of three cases of two fences $\left\|\textbf{PQ}\right\|$ and $\left\|\textbf{P'Q'}\right\|$ where they share no chains in (a), they share one chain $(\textbf{P}_{2}=\textbf{P'}_{2})$ in (b), and they share all chains $(\textbf{P}_{1}=\textbf{P'}_{1}, \textbf{P}_{2}=\textbf{P'}_{2},...,\textbf{P}_{n}=\textbf{P'}_{n})$ in (c).}
\label{fig:Figure9}
\end{figure}

Fences can have different configurations with respect to each other.  Two or more fences can share one chain, all chains or no chain as shown in Fig. \ref{fig:Figure9}.  We will show that \textit{if two fences share more than one chain, they share all chains}.  This is the analogue of the \textit{Parallel postulate} in the discrete form.     

\textit{Proof}:  Consider two fences $\left\|\textbf{P}_{1}\textbf{P}_{3}\right\|$ and $\left\|\textbf{P'}_{1}\textbf{P'}_{3}\right\|$.  If the fences share two chains $\textbf{P}_{1}$ and $\textbf{P'}_{1}$, $\textbf{P}_{2}$ and $\textbf{P'}_{2}$, then we have   

\begin{equation}
\label{eq:34}
\textbf{P}_{1}=\textbf{P'}_{1} \ \ , \ \  \textbf{P}_{2}=\textbf{P'}_{2}.
\end{equation}
Next, consider an event $x\in \textbf{P}_{1},\textbf{P'}_{1}$.  Event $x$ has to be collinear with both fences since it belongs to one of the chains in each fence, so we have 
\begin{equation}
\label{eq:35}
\textbf{P}_{3}\textbf{P}_{2}x \ \ , \ \  \overline{\textbf{P}_{3}\textbf{P}_{2}}x
\end{equation}
for the forward and backward projections of $x$, respectively.  Similarly for the projections of $x$ onto the fence $\left\|\textbf{P'}_{1}\textbf{P'}_{3}\right\|$ we have 
\begin{equation}
\label{eq:36}
\textbf{P'}_{3}\textbf{P'}_{2}x \ \ , \ \  \overline{\textbf{P'}_{3}\textbf{P'}_{2}}x.  
\end{equation}
Since $\textbf{P}_{2}=\textbf{P'}_{2}$, the projections given in Eq. (\ref{eq:35}) will become 
\begin{equation}
\label{eq:37}
\textbf{P}_{3}\textbf{P'}_{2}x \ \ \ \   \overline{\textbf{P}_{3}\textbf{P'}_{2}}x.
\end{equation}
Eqs. (\ref{eq:35}) and (\ref{eq:37}) can both be true only if $\textbf{P}_{3}\in \left\|\textbf{P'}_{1}\textbf{P'}_{3}\right\|$ which is only possible when $\textbf{P}_{3}=\textbf{P'}_{3}$, since all of the chains in the two fences are coordinated.

Note that the number of coordinated and collinear chains for each fence is arbitrary. So this can be generalized to a pair of fences with any number of chains.  Moreover, the number of chains in the two fences does not need to be equal.

This illustrates an analog of the Parallel postulate where fences that share no chain are called \textit{parallel}.  This result is not a postulate in this picture, rather it is a special case \textit{derived} through coordination condition.

Quantification of an interval with respect to a fence is discussed next when the interval does not belong to the subspace formed by the fence.  We will show how this quantification is the analog of a discrete version of the dot product.

\section{The Dot Product}
We can introduce a more general method that defines the projection of any interval onto a subspace defined by a fence (a set of coordinated and collinear chains).  We will show how the result of such projection is the analog of the dot product in the discrete form in a partially-ordered set of events.  

Consider a set of coordinated chains $\textbf{P}$, ..., $\textbf{Q}$ such that they form a fence $\left\| \textbf{PQ} \right\|$.  Consider also an event $x\notin \left\| \textbf{PQ} \right\|$.  The quantification of $x$ with respect to chain $\textbf{P}$ would be given by the pair $(Px,\overline{P}x)$.  Consider an event $p\in \textbf{P}$.  The quantification pair for the interval $[x,p]$ with respect to chain $\textbf{P}$ is given by $(p-Px,\overline{P}p-\overline{P}x)$.  The distance between event $x$ and chain $\textbf{P}$ which is the antisymmeric component of the symmetric-antisymmetric decomposition of the quantification pair, denoted $D(x,\textbf{P})$, is given as
\begin{equation}
\label{eq:38}
D(x,\textbf{P})=\frac{(p-Px)-(\overline{P}p-\overline{P}x)}{2}.
\end{equation}
Now consider another event $y$ such that $x\neq y$ and $y\notin \left\| \textbf{PQ} \right\|$.  Event $y$ is quantified by the pair $(Py,\overline{P}y)$ with respect to chain $\textbf{P}$.  The quantification of the interval $[y,p]$ with respect to chain $\textbf{P}$ is given by $(p-Py,\overline{P}p-\overline{P}y)$.  This gives the distance between event $y$ and chain $\textbf{P}$ as the antisymmetric component of the symmetric-antisymmetric decomposition of this quantification pair, denoted $D(y,\textbf{P})$ and defined as 
\begin{equation}
\label{eq:39}
D(y,\textbf{P})=\frac{(p-Py)-(\overline{P}p-\overline{P}y)}{2}.
\end{equation}   
Similarly, we can have the distance between events $x$ and $y$ and chain $\textbf{Q}$ respectively, as
\begin{eqnarray}
\label{eq:40}
D(x,\textbf{Q})=\frac{(q-Qx)-(\overline{Q}q-\overline{Q}x)}{2} \\
D(y,\textbf{Q})=\frac{(q-Qy)-(\overline{Q}q-\overline{Q}y)}{2}. 
\end{eqnarray}
The projection of the interval $[x,y]$ onto the subspace $\left\langle \textbf{PQ} \right\rangle$ will then be given by 
\begin{equation}
\label{eq:42}
(D(y,\textbf{P})^{2}-D(x,\textbf{P})^{2})-(D(y,\textbf{Q})^{2}-D(x,\textbf{Q})^{2}),
\end{equation} 
or equivalently
\begin{equation}
\label{eq:43}
(D(y,\textbf{P})^{2}-D(y,\textbf{Q})^{2})-(D(x,\textbf{P})^{2}-D(x,\textbf{Q})^{2}).
\end{equation}
Normalizing this term by twice the distance between $\textbf{P}$ and $\textbf{Q}$, $2D(\textbf{P},\textbf{Q})$, gives
\begin{equation}
\label{eq:44}
\frac{(D(y,\textbf{P})^{2}-D(y,\textbf{Q})^{2})-(D(x,\textbf{P})^{2}-D(x,\textbf{Q})^{2})}{2D(\textbf{P},\textbf{Q})}.
\end{equation}

\begin{figure}[t]
\centering
\includegraphics[width=0.50
\textwidth]{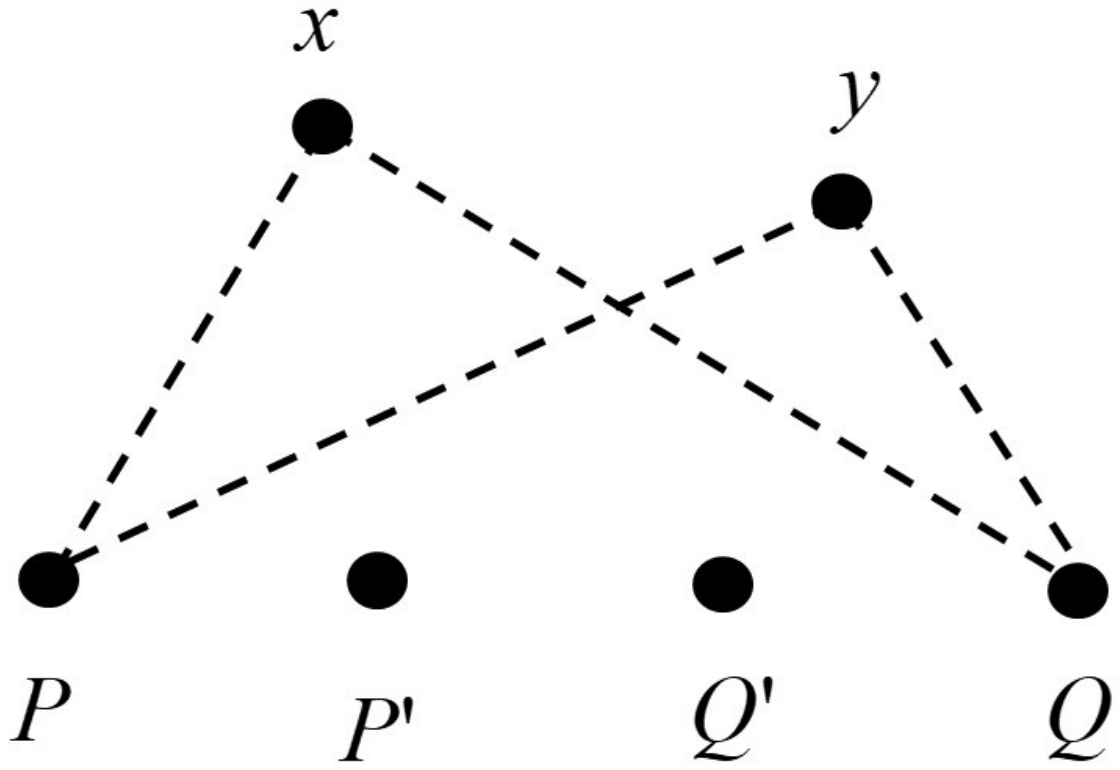}  
\caption{Geometric view of interval $[x,y]$ projected onto fence $\left\|\textbf{PQ}\right\|$.}
\label{fig:Figure10}
\end{figure}

This is a method of subspace projection, and below we will show that Eq. (\ref{eq:44}) is the analog of the dot product for the poset picture.  Note that since this relation only depends on the antisymmetric components, it is independent of the choice of coordinated chains as long as they belong to the same 1+1 dimensional subspace shown in Fig. \ref{fig:Figure10}.

To prove that Eq. (\ref{eq:44}) is the answer, we seek a function $f$ that gives us the projection of $[x,y]$ onto $\left\|\textbf{PQ}\right\|$ such that the function is independent of the choice of quantifying chains in the $\left\|\textbf{PQ}\right\|$ fence.  To satisfy this requirement, the function must result in a scalar and it must depend on the distances between events $x$ and $y$ and chains $\textbf{P}$ and $\textbf{Q}$.  So we can assume that the function should linearly depend on all possible scalars involved in this projection  

\begin{equation}
\label{eq:45}
f(D(x,\textbf{P})^2,D(x,\textbf{Q})^2,D(y,\textbf{P})^2,D(y,\textbf{Q})^2) \\ 
=aD(x,\textbf{P})^2+bD(x,\textbf{Q})^2+cD(y,\textbf{P})^2+dD(y,\textbf{Q})^2, 
\end{equation}
where the distance $D(x,\textbf{P})$ between event $x$ and chain $\textbf{P}$ has the scalar representation of $D(x,\textbf{P})^{2}=-\left((p_{x}-\bar{p}_x)/2 \right)^{2}$.  Similarly, the scalars $D(x,\textbf{Q})^2$, $D(y,\textbf{P})^2$ and $D(y,\textbf{Q})^2$ are found from  
\begin{equation*}
D(x,\textbf{Q})=\frac{q_{x}-\bar{q}_{x}}{2},\ \ \ \ D(y,\textbf{P})=\frac{p_{y}-\bar{p}_{y}}{2},\ \ \ \ D(y,\textbf{Q})=\frac{q_{y}-\bar{q}_{y}}{2}.
\end{equation*}  
 
Note that instead of $\textbf{P}$ and $\textbf{Q}$, we can consider any other chain from the fence and the result would be the same.  The linear combination of the distances ensures that the result of the projection remains a scalar of the same order.  Since $[x,y]$ has the same time coordinate with respect to all chains in the fence, there is no length-dependence in the projection function.  

We now need to determine constants $a$, $b$, $c$ and $d$.  Consider the following special cases:

\textit{Special Case I}:  If $x=y$ then $[x,y]=0$.  So the projection of $[x,y]$ onto $\left\|\textbf{PQ}\right\|$ will be
\begin{equation}
\label{eq:46}
aD(x,\textbf{P})^2+bD(x,\textbf{Q})^2+cD(y,\textbf{P})^2+dD(y,\textbf{Q})^2=0,
\end{equation} 
which gives
\begin{equation}
\label{eq:47}
a=-c\ \ \ \ \ \ \ b=-d.
\end{equation}
Substituting these results back into Eq. (\ref{eq:46}) gives
\begin{equation}
\label{eq:48}
aD(x,\textbf{P})^2+bD(x,\textbf{Q})^2-aD(y,\textbf{P})^2-bD(y,\textbf{Q})^2=0.
\end{equation}

\begin{figure}[t]
\centering
\includegraphics[width=0.30
\textwidth]{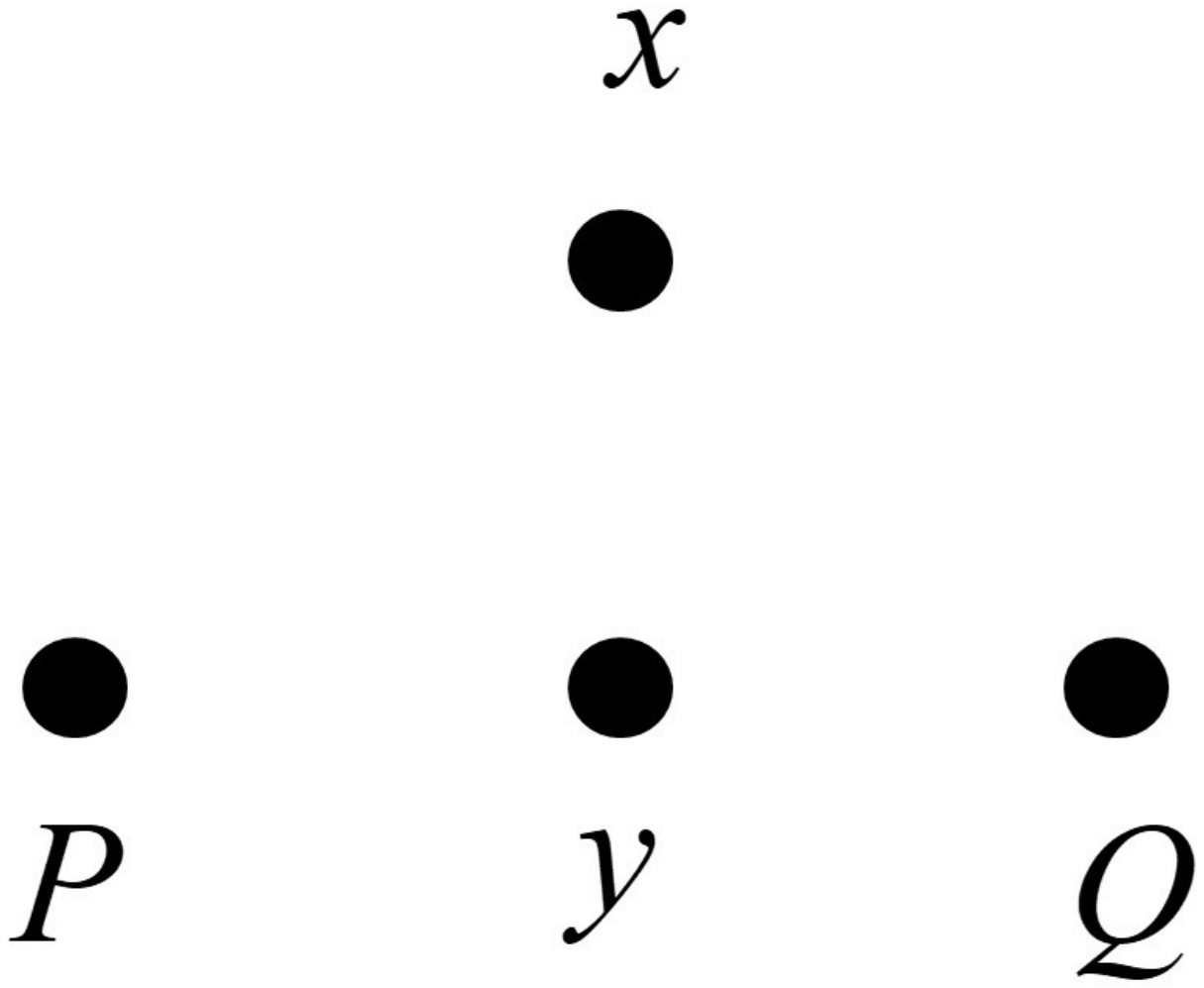}  
\caption{Geometric view of interval $[x,y]$ perpendicular to fence $\left\|\textbf{PQ}\right\|$.}
\label{fig:Figure11}
\end{figure}

\textit{Special Case II}:  Consider a special configuration where $y\in \left\|\textbf{PQ}\right\|$ but $y\notin\textbf{P},\textbf{Q}$, and $x\notin\left\|\textbf{PQ}\right\|$.  Also for their distances we have $D(x,\textbf{P})=D(x,\textbf{Q})$ and $D(y,\textbf{P})=D(y,\textbf{Q})$, and the time coordinate for both $x$ and $y$ is the same as quantified by the chains in the fence, as shown in Fig. \ref{fig:Figure11}.  We can write the Pythagorean theorem for this configuration as

\begin{equation}
\label{eq:49} 
D(y,\textbf{P})^2-D(x,\textbf{P})^2=D(y,\textbf{Q})^2-D(x,\textbf{Q})^2
\end{equation}
in which case the projection will also be zero.  Using this condition in Eq. (\ref{eq:48}) gives
\begin{equation}
\label{eq:50}
a=-b.
\end{equation}

\textit{Special Case III}:  The last special case that we consider is when $x\in \textbf{P}$ and $y\in \textbf{Q}$.  The projection of $[x,y]$ onto $\left\| \textbf{P}\textbf{Q} \right\|$ would be $D(\textbf{P},\textbf{Q})$.  So we have
\begin{equation}
\label{eq:51}
a(D(x,\textbf{P})^2-D(y,\textbf{P})^2-D(x,\textbf{Q})^2+D(y,\textbf{Q})^2)=D(\textbf{P},\textbf{Q}).
\end{equation}
However, if $x\in \textbf{P}$ and $y\in \textbf{Q}$ then
\begin{eqnarray*}
D(x,\textbf{P})^2=0 \\ D(y,\textbf{Q})^2=0 \\ D(x,\textbf{Q})^2=D(y,\textbf{P})^2=D(\textbf{P},\textbf{Q})^2.
\end{eqnarray*}  
Substituting these results back into Eq. (\ref{eq:51}) gives
\begin{equation}
\label{eq:52}
a(D(\textbf{P},\textbf{Q})^2+D(\textbf{P},\textbf{Q})^2)=D(\textbf{P},\textbf{Q}),
\end{equation}
which results in getting the value for $a$ as
\begin{equation}
\label{eq:53}
a=\frac{1}{2D(\textbf{P},\textbf{Q})}.
\end{equation}

Thus the projection of an interval $[x,y]$ onto a set of coordinated and collinear chains that form a fence $\left\| \textbf{P}\textbf{Q} \right\|$ where both $x$ and $y$ are quantified with the same time coordinates with respect to the chains in the fence $\left\|\textbf{PQ}\right\|$ is given by
\begin{equation}
\label{eq:54}
\left|\frac{D(y,\textbf{P})^2-D(x,\textbf{P})^2-D(y,\textbf{Q})^2+D(x,\textbf{Q})^2}{2D(\textbf{P},\textbf{Q})}\right|\equiv D(x',y').
\end{equation} 

Since all chains in the fence are coordinated they can share quantifications of the same event.  Thus the same proof holds for any other chain in the fence as it does for $\textbf{P}$ and $\textbf{Q}$.  This allows us to write the discrete form of the dot product of an interval $[x,y]$ and a fence $\left\|\textbf{PQ}\right\|$ as
\begin{equation}
\label{eq:55}
D(x,y).D(\textbf{P},\textbf{Q})\equiv \left|\frac{D(y,\textbf{P})^2-D(x,\textbf{P})^2-D(y,\textbf{Q})^2+D(x,\textbf{Q})^2}{2}\right|.
\end{equation}

Finally, we can extend quantification with a fence to quantification with a number of fences.  We will study a special configuration of fences, called a \textit{grid}, and show how such quantifications result in a discrete version of the geometric product.

\section{Grid}
Let us now study the case where we have more than one fence.  In this section we will focus on a specific configuration of fences called a \textit{grid} and we will look at the projection of an interval onto the subspace formed by the grid.  

\begin{figure}[t]
\centering
\includegraphics[width=0.70
\textwidth]{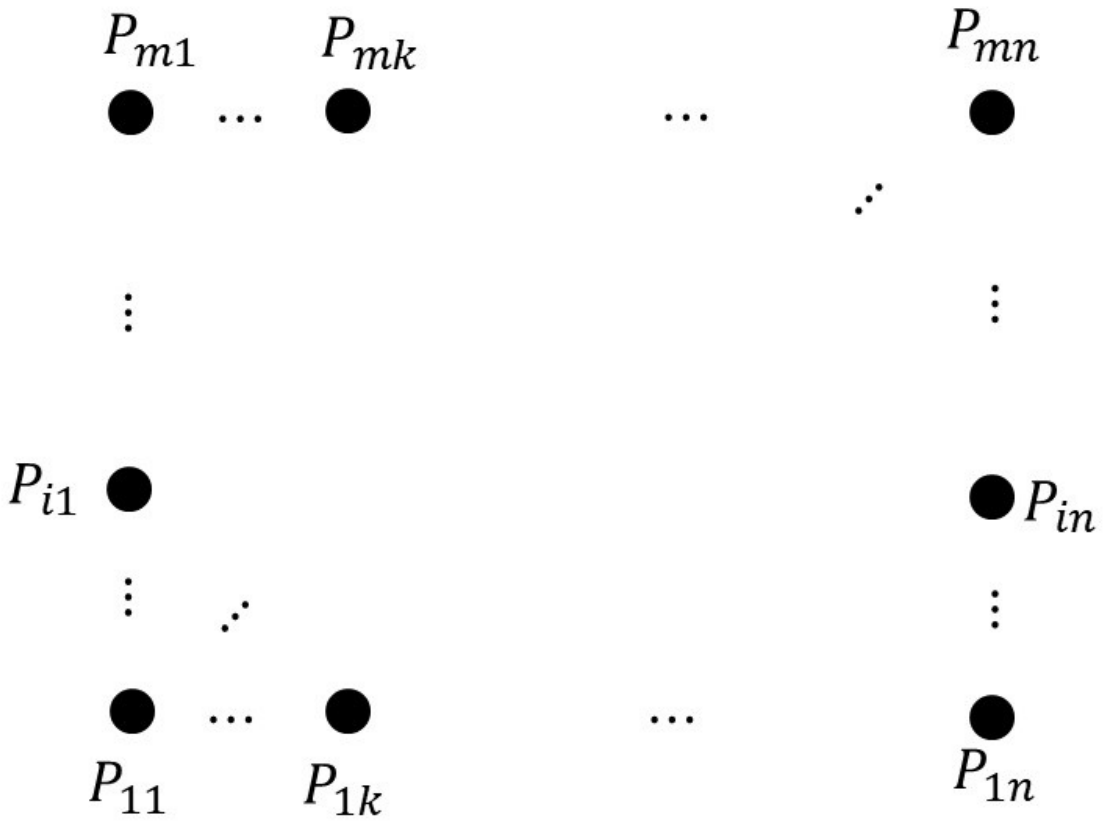}  
\caption{Geometric view of the grid $\diamond \textbf{P}_{11}\textbf{P}_{mn}$.  Fences $\left\| \textbf{P}_{11}\textbf{P}_{1n} \right\|$, ..., $\left\| \textbf{P}_{m1}\textbf{P}_{mn} \right\|$ are parallel and fences  $\left\| \textbf{P}_{11}\textbf{P}_{m1} \right\|$, $\left\| \textbf{P}_{12}\textbf{P}_{m2} \right\|$, ..., $\left\| \textbf{P}_{1n}\textbf{P}_{mn} \right\|$ are also parallel.}
\label{fig:Figure12}
\end{figure}

A \textit{grid}, denoted $\diamond \textbf{P}_{11}\textbf{P}_{mn}$, is defined to be a set of two or more parallel fences, $\left\| \textbf{P}_{11}\textbf{P}_{1n} \right\|, ..., \left\| \textbf{P}_{m1}\textbf{P}_{mn}\right\|$ where $m,n\in \left\{1,2,3,...\right\}$, such that $\left\| \textbf{P}_{11}\textbf{P}_{m1} \right\|$, $\left\| \textbf{P}_{12}\textbf{P}_{m2} \right\|$, ..., $\left\| \textbf{P}_{1n}\textbf{P}_{mn} \right\|$ also form parallel fences as shown in Fig. (\ref{fig:Figure12}).  Thus the subspace formed by a grid is 2+1.  The two indices in each chain indicate fence and chain numbers, respectively.  For example, $\textbf{P}_{23}$ is the third chain of the second fence in the grid.

\subsection{Quantification Inside a Grid}
\begin{figure}[t]
\centering
\includegraphics[width=0.70
\textwidth]{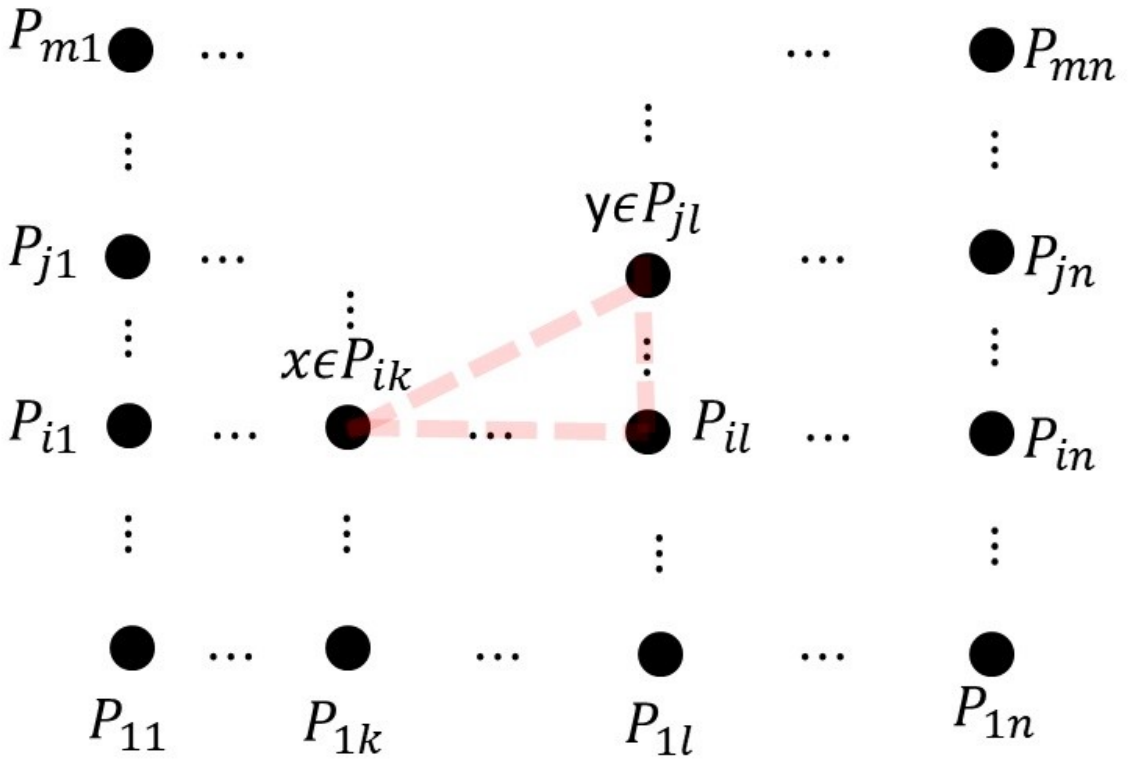}  
\caption{Events $x, y$ are on $\textbf{P}_{ik}\in \left\|\textbf{P}_{i1},\textbf{P}_{in} \right\|$ and $\textbf{P}_{jl}\in \left\|\textbf{P}_{j1}\textbf{P}_{jn} \right\|$, respectively.}
\label{fig:Figure13}
\end{figure}

Consider events $x,y$ such that $x\in \left\| \textbf{P}_{i1}\textbf{P}_{in} \right\|$ and $y\in \left\| \textbf{P}_{j1}\textbf{P}_{jn} \right\|$, where $\left\| \textbf{P}_{i1}\textbf{P}_{in} \right\|, \left\| \textbf{P}_{j1}\textbf{P}_{jn} \right\|\in \diamond \textbf{P}_{11}\textbf{P}_{mn}$ are arbitrary fences in the grid shown in Fig. (\ref{fig:Figure13}).

Next, we use Eq. (\ref{eq:54}) to write the projection of interval $[x,y]\in \diamond \textbf{P}_{11}\textbf{P}_{mn}$ onto the fence $\left\|\textbf{P}_{11}\textbf{P}_{1n} \right\|$  
\begin{equation}
\label{eq:56}
\left|\frac{D(y,\textbf{P}_{11})^2-D(x,\textbf{P}_{11})^2-D(y,\textbf{P}_{1n})+D(x,\textbf{P}_{1n})^2}{2D(\textbf{P}_{11},\textbf{P}_{1n})} \right|\equiv D(\textbf{P}_{1k},\textbf{P}_{1l}) .
\end{equation}
Note that the projection can be written using any chain in the fence.  Moreover, since fences are coordinated, one can pick any other fence and find the projection of the interval $[x,y]$ onto that fence using Eq. (\ref{eq:54}).  For instance, using $\left\|\textbf{P}_{i1}\textbf{P}_{in} \right\|$, the projection would be given by $D(\textbf{P}_{ik},\textbf{P}_{il})$. 

If the relation between the projections of any two consecutive fences, such as $\left\|\textbf{P}_{i1}\textbf{P}_{in} \right\|$ and $\left\|\textbf{P}_{j1}\textbf{P}_{jn} \right\|$, is such that one can write the Pythagorean theorem as
\begin{equation}
\label{eq:57}
D(\textbf{P}_{ik},\textbf{P}_{jl})^2=D(\textbf{P}_{ik},\textbf{P}_{il})^2+D(\textbf{P}_{jl},\textbf{P}_{il})^2,
\end{equation}
where $D(\textbf{P}_{ik},\textbf{P}_{jl})\equiv D(x,y)$ is the distance between the two chains $\textbf{P}_{ik}$ and $\textbf{P}_{jl}$, then the grid is defined to be \textit{orthogonal}. 
If the grid is orthogonal, Eq. (\ref{eq:56}) can be used to rewrite Eq. (\ref{eq:57}) as
\begin{multline}
\label{eq:58}
D(x,y)^2=\left(\frac{D(y,\textbf{P}_{11})^2-D(x,\textbf{P}_{11})^2-D(y,\textbf{P}_{1n})+D(x,\textbf{P}_{1n})^2}{2D(\textbf{P}_{11},\textbf{P}_{1n})} \right)^2\\
+D(\textbf{P}_{jl},\textbf{P}_{il})^2. 
\end{multline}
Rearranging this gives
\begin{multline}
\label{eq:59}
D(x,y)^2D(\textbf{P}_{11},\textbf{P}_{1n})^2=\left(\frac{D(y,\textbf{P}_{11})^2-D(x,\textbf{P}_{11})^2-D(y,\textbf{P}_{1n})+D(x,\textbf{P}_{1n})^2}{2} \right)^{2}\\ 
+D(\textbf{P}_{jl},\textbf{P}_{il})^2 D(\textbf{P}_{11},\textbf{P}_{1n})^2.
\end{multline}
The first term on the right hand side is simply the dot product.  So Eq. (\ref{eq:59}) can be rewritten as
\begin{multline}
\label{eq:60}
D(x,y)^2 D(\textbf{P}_{11},\textbf{P}_{1n})^2 = (D(x,y). D(\textbf{P}_{11},\textbf{P}_{1n}))^2\\
+ (D(y,\textbf{P}_{1l})-D(\textbf{P}_{il},\textbf{P}_{1l}))^2 D(\textbf{P}_{11},\textbf{P}_{1n})^2,
\end{multline}
where we substituted $D(y,\textbf{P}_{1l})-D(\textbf{P}_{il},\textbf{P}_{1l})$ for $D(y,\textbf{P}_{il})$.
To study the second term on the right hand side, we can look at a few special cases:

\textit{Special Case I}: If $x\in \textbf{P}_{ik}$ and $y\in \textbf{P}_{il}$, then $D(y,P_{il})=0$.

\textit{Special Case II}: If $x\in \textbf{P}_{il}$ and $y\in \textbf{P}_{jl}$, then $(D(y,P_{il})-0)D(\textbf{P}_{ik},\textbf{P}_{il})=D(x,y)D(\textbf{P}_{ik},\textbf{P}{il})$.

\textit{Special Case III}: If $x\in \textbf{P}_{jl}$ and $y\in \textbf{P}_{il}$, then $(0-D(x,P_{il}))D(\textbf{P}_{ik},\textbf{P}_{il})= -D(x,y)D(\textbf{P}_{ik},\textbf{P}_{il})$.

Since $D(x,y)$ is the distance between two events $x$ and $y$ or equivalently, the distance between their corresponding chains $\textbf{P}_{ik}$ and $\textbf{P}_{jl}$, and $D(\textbf{P}_{ik},\textbf{P}_{il})$ is the distance between two chains $\textbf{P}_{ik}$ and $\textbf{P}_{il}$, their product must be associated with an \textit{area}, and the negative sign in the third special case indicates that this must be a \textit{directed} area.  The three special cases considered here indicate that this term can be regarded as the analog of the outer (wedge) product in a discrete case. So Eq. (\ref{eq:60}) can be rewritten as 
\begin{equation}
\label{eq:61}
D(x,y)^2 D(\textbf{P}_{ik},\textbf{P}_{il})^2=(D(x,y) . D(\textbf{P}_{ik},\textbf{P}_{il}))^2+(D(x,y) \wedge D(\textbf{P}_{ik},\textbf{P}_{il}))^2,
\end{equation}
where $D(\textbf{P}_{1k},\textbf{P}_{1l})$ is substituted by $D(\textbf{P}_{ik},\textbf{P}_{il})$.  

The same result can be found in the geometric product of two vectors $\vec{a},\vec{b}$ which is defined as the sum of their inner and their outer products
\begin{equation}
\label{eq:62}
\vec{a}\vec{b}=\vec{a}.\vec{b} + \vec{a}\wedge \vec{b}.
\end{equation}  
Since $\vec{a}\wedge \vec{b} = -\vec{b}\wedge \vec{a}$, we can write
\begin{equation}
\label{eq:63}
\vec{a}.\vec{b}=(\vec{a}\vec{b}+\vec{b}\vec{a})/2\ ,\ \ \ \vec{a}\wedge \vec{b}=(\vec{a}\vec{b}-\vec{b}\vec{a})/2.
\end{equation}
Squaring both equations and adding them gives
\begin{equation}
\label{eq:64}
(\vec{a}\vec{b})^2=(\vec{a}.\vec{b})^2+(\vec{a}\wedge \vec{b})^2,
\end{equation}
which is the same result found in Eq. (\ref{eq:61}).

The wedge product found here is based on considering 2+1 dimensions.  However, the results can be generalized to higher dimensions $n$ where $n\in N$.  In the special case where $n=3$ we have 3+1 dimensions where the wedge product can be written as the cross product which gives Eq.(64) as $(\vec{a}\vec{b})^2=(\vec{a}.\vec{b})^2+(\vec{a}\times \vec{b})^2$ .

\section{Conclusion}
In this paper, we applied the quantification technique introduced in Ref. \cite{KB1} for quantifying partially-ordered set of events to different configurations of chains in order to find geometric features of a poset of events. This lead to a discrete version of some features of Euclidean
geometry. We first reviewed our previous work on quantifying a partially-ordered set of events using chain projections. We showed how order in projections of an event onto a pair of chains induce some of the fundamental concepts in geometry such as collinearity, directionality and subspaces. Coordination condition from our previous work is reviewed and then extended from two to more chains.

Our results suggest that many features in Euclidean geometry are not fundamental, but rather they are derived from order. Our main findings in this paper can be outlined as follows:

\begin{description}
\item[{[1]}] We demonstrated that the Pythagorean theorem is derived from the fact that the interval scalars of orthogonal subspaces are additive (see Eq. (\ref{eq:10})).

\item[{[2]}] We showed that geometric shapes are formed from the quantification of more than two equidistant coordinated chains which gives rise to multiple spatial dimensions (see Eqs. (\ref{eq:24}), (\ref{eq:33})).

\item[{[3]}] We introduced the concept of a fence as a set of three or more collinear and coordinated chains. Then, we studied different configurations of two fences. Most importantly, we proved (see Eqs. (\ref{eq:35}-\ref{eq:37})) that fences that share more than one chain, share all chains which is the equivalent of the Parallel postulate in the discrete case.

\item[{[4]}] We found that the projection of an interval onto a set of collinear and coordinated chains results in the dot product (see Eq. (\ref{eq:54})).  The features of the outer (wedge) product in $2+1$ dimensions appeared when quantification was extended from one fence to a number of fences.

\item[{[5]}] Writing the Pythagorean theorem inside a grid, we found a relation whose terms were similar to those of the geometric product squared (see Eq. (\ref{eq:61})).
\end{description}

Applications of lattice theory extend beyond Euclidean geometry.  For instance, lattice theory has a wide range of applications from computer science (e.g. \cite{Fayyad}\cite{MR}) to physics (e.g. \cite{SMphys}).  Applications of causal sets have also been investigated.  In one study, the dynamics of spacetime at quantum scales was studied using numerical simulations of a two-dimensional causal set quantum gravity \cite{Su2}.  These results indicate that changes in some parameters cause a change in the properties of the causal set, which is referred to as a phase transition.  Moreover, there have been studies on testing the dimensionality of spacetime using a causal set model where it has been demonstrated that the dimensions of spacetime, for conformally flat spacetime, can be determined by the causal set theory \cite{Reid}.  The approach is based on embedding the causal set in conformally flat spacetime which can ultimately result in the correct continuum dimensions.

Deriving the geometry of spacetime has been the focus of many studies.  For instance, in order to study curved spacetime, a topology which unifies causal, differential and conformal structure was introduced \cite{Hawking}.  In another example, the topology of spacetime was shown to be determined by a class of timelike curves which are all possible causal trajectories of particles with mass \cite{Malament}.  Another study models causal sets that correspond to continuum spacetime and compares their discrete topology to that of the continuum topology \cite{Su3}.   

It is important to emphasize that the results found in our paper are based on no assumptions about the structure or even the existence of space or time.  Rather, they appear merely as a result of order in a poset of events and their consistent quantification.  The results found so far in this picture indicate an interesting view of the foundations of physics without any assumptions about space or time.  In particular, this theoretical approach leads to obtaining many features of the Fermion physics and the Dirac equation based on a model of a free electron as a poset \cite{Kquant}\cite{Kcontemp}\cite{Kelectron}\cite{KInfo2}, relativistic Newton's second law when a particle in being influenced \cite{KW}, and the derivation of all aspects of Special Relativity including the Minkowski metric, the existence of an upper limit for velocity, and the Lorentz transformations \cite{KB1}\cite{KB2}.  Finally, this methodology may be used to investigate further physical laws and their connections based on a minimalist view.  Based on our previous work \cite{KB1}\cite{KB2} the scalar quantification of symmetric and antisymmetric pairs associated with quantifications along and across a pair of coordinated chains, respectively, are responsible for the Lorentzian signature that appears in this picture.  However, nothing in the causal relation or the quantification scheme we have used prevents us from going to N spatial dimensions.  So, the fact that the laws of physics, expressed in terms of empirically defined concepts such as mass, energy, momentum, and positions in space and time, emerge from a more foundational and basic axiomatic/theoretical approach suggests that these concepts may not be fundamental.  Based on this picture, one could argue that what we observe in space-time physics is a macroscopic manifestation of a continuum of elementary events which, when suitably ordered,  can be expressed as time, compared in terms of distances, and grouped in certain ways to form areas and/or volumes while there is no space or time defined in this picture.  Indeed, it happens that one can argue that this macroscopic physical picture can be regarded, at a deeper level, as emerging from a microscopic discrete theoretical framework where the concepts of a causal set equipped with an ordering relation can give rise to a classical space-time equipped with its Lorentzian metric. This foundational viewpoint lays the foundations of the so-called causal set theory approach to quantum gravity \cite{Bombelli1}\cite{Su}.

Although the relative view of space as being merely a relation among objects which was envisioned by al-Ghazali \cite{Ghazali} and later Leibnitz \cite{LB} was first ignored and then rejected due to its contradiction with the Newtonian mechanics, it was not until the derivation of Einstein's theory of Special Relativity when the relative view of space and time was taken seriously.  The poset picture that was discussed in this research takes this a step further by showing how all aspects of Special Relativity can be derived with no assumption about space and time \cite{KB1}.  Even geometrical properties that were thought to be intrinsic properties of space such as directionality and subspaces emerge in this picture with no reference to space and time.  As Poincar\'{e} and Mach also suggested, it is all about the connectivity.

\section*{Acknowledgements}
N.B. thanks W. Ballard, D. McCuskey, C. Zhang and I. Ashley for helpful discussions on deriving some of aspects in Euclidean geometry including the dot product, the Parallel postulate and the outer (wedge) product.  The authors thank the anonymous referees for useful comments leading to an improved version of the manuscript.

\section{Appendix}
As mentioned in section 3.1 there are five possible cases where an event $x$ and two distinct chains $\textbf{P}$ and $\textbf{Q}$ in a partially-ordered set can be collinear.  The proof is given here. 

\textbf{Theorem:}  According to the definition of collinearity the following cases exhaust all possible ways in which $x$ can be collinear with $\textbf{P}$ and $\textbf{Q}$:
   
Case I:
\begin{equation*}
Px=\overline{P}Qx  \ \ \ \ \ \ \  Qx= QPx 
\end{equation*}
\begin{equation*}
\overline{P}x=P\overline{Q}x  \ \ \ \ \ \ \  \overline{Q}x=\overline{Q}\overline{P}x
\end{equation*}

Case II:
\begin{equation*}
Px=P\overline{Q}x  \ \ \ \ \ \ \  Qx= Q\overline{P}x 
\end{equation*}
\begin{equation*}
\overline{P}x=\overline{P}Qx  \ \ \ \ \ \ \  \overline{Q}x=\overline{Q}Px
\end{equation*}

Case III:
\begin{equation*}
Px=PQx  \ \ \ \ \ \ \  Qx= \overline{Q}Px 
\end{equation*}
\begin{equation*}
\overline{P}x=\overline{P}\overline{Q}x  \ \ \ \ \ \ \  \overline{Q}x=Q\overline{P}x
\end{equation*}

Case IV:
\begin{equation*}
Px=PQx  \ \ \ \ \ \ \  Qx= \overline{Q}Px 
\end{equation*}
\begin{equation*}
\overline{P}x=P\overline{Q}x  \ \ \ \ \ \ \  \overline{Q}x=\overline{Q}\overline{P}x
\end{equation*}

Case V:
\begin{equation*}
Px=\overline{P}Qx  \ \ \ \ \ \ \  Qx= QPx 
\end{equation*}
\begin{equation*}
\overline{P}x=\overline{P}\overline{Q}x  \ \ \ \ \ \ \  \overline{Q}x=Q\overline{P}x.
\end{equation*}

\textit{Proof.}  Consider two distinct finite chains $\textbf{P}\in \Pi$ and $\textbf{Q}\in \Pi$ and an event $x\in \Pi$ such that event $x$ is forward and backward projected onto both chains.  There are $4^{2}=16$ algebraic possibilities of the order of projections of $x$ onto the two chains would be given by 
\begin{equation}
\begin{array}{cccc} Px=PQx & Px=P\overline{Q}x & Px=\overline{P}Qx & Px=\overline{P}\overline{Q}x \\ \overline{P}x=\overline{P}Qx & \overline{P}x=\overline{P}\overline{Q}x & \overline{P}x=P\overline{Q}x & \overline{P}x=PQx \\ Qx=QPx & Qx=Q\overline{P}x & Qx=\overline{Q}Px & Qx=\overline{Q}\overline{P}x \\ \overline{Q}x=\overline{Q}Px & \overline{Q}x=\overline{Q}\overline{P}x & \overline{Q}x=Q\overline{P}x & \overline{Q}x=QPx.
\end{array}
\end{equation}
There are then $(4^{2})^{2}=256$ pairwise combinations of these relations.  First note that cases $Px=\overline{P}\overline{Q}x$, $\overline{P}x=PQx$, $Qx=\overline{Q}\overline{P}x$ and $\overline{Q}x=QPx$ cannot occur.  Consider, for example $Px=\overline{P}\overline{Q}x$.  On the right hand side we have both backward projections $\overline{P}$ and $\overline{Q}$ of event $x$ whereas in the left hand side we have only forward projection $P$ of the same event on one of the chains.  This is not possible since if $x$ is backward projected onto chains $\textbf{P}$ and $\textbf{Q}$ then $\overline{Q}x>\overline{P}x$ which is only equal to $Px$ if $Px=\overline{P}x$ which violates causality.  Similarly the other three cases $\overline{P}x=PQx$, $Qx=\overline{Q}\overline{P}x$ and $\overline{Q}x=QPx$ are also ruled out.  This leaves us with twelve cases 
\begin{equation}
\begin{array}{ccc} 0 & 1 & 2 \\ Px=PQx & Px=P\overline{Q}x & Px=\overline{P}Qx \\ \overline{P}x=\overline{P}Qx & \overline{P}x=\overline{P}\overline{Q}x & \overline{P}x=P\overline{Q}x \\ Qx=QPx & Qx=Q\overline{P}x & Qx=\overline{Q}Px \\ \overline{Q}x=\overline{Q}Px & \overline{Q}x=\overline{Q}\overline{P}x & \overline{Q}x=Q\overline{P}x 
\end{array}
\end{equation}
where the columns are codes as 0, 1 and 2.  We have to consider all the possible combinations of the forward and backward projections for both chains, which is $P$, $Q$, $\overline{P}$ and $\overline{Q}$ for each combination, which in terms of the codes given above, would be
\begin{eqnarray*}
0000 \\ 0001 \\ 0002 \\ \vdots \\ 2222  
\end{eqnarray*}
which are $3^{4}=81$ cases.  We now have to consider each case to see which of them are possible.

Consider the case 0000
\begin{equation}
\begin{array}{cccc} Px=PQx & \overline{P}x=\overline{P}Qx & Qx=QPx & \overline{Q}x=\overline{Q}Px.
\end{array}
\end{equation}
These relations among the forward and backward projections of event $x$ onto the two distinct chains $\textbf{P}$ and $\textbf{Q}$ are also not possible.  For example, if we have $Px=PQx$, it means that the forward projection of event $x$ onto chain $\textbf{P}$ is first found on chain $\textbf{Q}$ while $Qx=QPx$ indicates that the forward projection of event $x$ onto chain $\textbf{Q}$ is first found on chain $\textbf{P}$.  This is only possible when chains $\textbf{P}$ and $\textbf{Q}$ are interchanged which then does not refer to this one case.  This also holds for the backward projections, that is, the two cases $\overline{P}x=\overline{P}\overline{Q}x$ and $\overline{Q}x=\overline{Q}\overline{P}x$ cannot occur together.  There are 17 combinations that include these two conditions together which will be eliminated.
Next consider 0010     
\begin{equation}
\begin{array}{cccc} Px=PQx & \overline{P}x=\overline{P}Qx & Qx=Q\overline{P}x & \overline{Q}x=\overline{Q}Px.
\end{array}
\end{equation}   
Substituting $Qx=Q\overline{P}x$ into $Px=PQx$ gives
\begin{equation}
Px=PQ\overline{P}x
\end{equation}
which is impossible since on the right hand side forward and backward projections of $x$ onto $\textbf{P}$ overlap.  Similarly, $Px=P\overline{Q}x$ and $Qx=QPx$ cannot be combined which together eliminates 14 more cases that include these combinations.  For the same reason we can see that the two cases $\overline{P}x=P\overline{Q}x$ and $\overline{Q}x=\overline{Q}Px$ as well as $\overline{Q}x=Q\overline{P}x$ and $\overline{P}x=\overline{P}Qx$ also cannot be combined which eliminates 13 more cases.  Moreover, the two cases $Px=\overline{P}Qx$ and $\overline{P}x=\overline{P}Qx$ cannot be combined since they yield that both $Px$ and $\overline{P}x$ are equal to $\overline{P}Qx$.  Therefore 8 more cases are ruled out.  
Now consider $\overline{P}x=\overline{P}\overline{Q}x$ and $\overline{Q}x=\overline{Q}Px$.  Upon substitution of $\overline{Q}x$ from the second equation into the first we obtain $\overline{P}x=\overline{P}\overline{Q}Px$ which is an impossible order of projections since the forward projection of event $x$ onto $\textbf{P}$ overlaps with its backward projection.  Similarly, the combinations $\overline{P}x=\overline{P}Qx$ and $\overline{Q}x=\overline{Q}\overline{P}x$ cannot occur either which together, eliminates 9 more cases.  The two cases $Qx=Q\overline{P}x$ and $\overline{Q}x=Q\overline{P}x$ cannot be combined since they give the same relation for the forward and backward projections.  The same situation holds for $Px=P\overline{Q}x$ and $\overline{P}x=P\overline{Q}x$.  These two conditions rule out 5 more cases.  It is also impossible to have $Px=P\overline{Q}x$ and $\overline{P}x=\overline{P}\overline{Q}x$ together as well as $Qx=Q\overline{P}x$ and $\overline{Q}x=\overline{Q}\overline{P}x$ since both forward and backward projections of $x$ onto one of the two chains are first found as a backward projection of $x$ onto the other chain which contradicts collinearity.  This eliminates 3 more cases.  Consider the combinations that has $Px=\overline{P}Qx$ and $Qx=\overline{Q}Px$.  If we substitute the value for $Qx$ from the second equation into the first one we have $Px=\overline{P}\overline{Q}Px$ which is impossible, which eliminates 2 more cases.  The two $Px=PQx$ and $\overline{P}x=\overline{P}Qx$ cannot be combined either since both forward and backward projections of $x$ onto $\textbf{P}$ can be found on the forward projection of $x$ onto $\textbf{Q}$ which eliminates 2 more cases.  The two cases $\overline{P}x=\overline{P}Qx$ and $Qx=\overline{Q}Px$ cannot be combined either since if we substitute $Qx$ into the relation for $\overline{P}x$ we get $\overline{P}x=\overline{P}\overline{Q}Px$ which means that the forward and backward projections of $x$ onto $\textbf{P}$ are the same point.  This eliminates 2 more cases.  Finally, the combination 0121 cannot occur which refers to $Px=PQx$, $\overline{P}x=\overline{P}\overline{Q}x$, $Qx=\overline{Q}Px$ and $\overline{Q}x=\overline{Q}\overline{P}x$.   If we have $Px=PQx$, $\overline{P}x=\overline{P}\overline{Q}x$, $\overline{Q}x=\overline{Q}\overline{P}x$, then we cannot have $Qx=\overline{Q}Px$.  This leaves us with the remaining five cases
 
Case I:
\begin{equation*}
Px=\overline{P}Qx  \ \ \ \ \ \ \  Qx= QPx 
\end{equation*}
\begin{equation*}
\overline{P}x=P\overline{Q}x  \ \ \ \ \ \ \  \overline{Q}x=\overline{Q}\overline{P}x
\end{equation*}
which corresponds to the code 2201

Case II:
\begin{equation*}
Px=P\overline{Q}x  \ \ \ \ \ \ \  Qx= Q\overline{P}x 
\end{equation*}
\begin{equation*}
\overline{P}x=\overline{P}Qx  \ \ \ \ \ \ \  \overline{Q}x=\overline{Q}Px
\end{equation*}
which corresponds to the code 1010

Case III:
\begin{equation*}
Px=PQx  \ \ \ \ \ \ \  Qx= \overline{Q}Px 
\end{equation*}
\begin{equation*}
\overline{P}x=\overline{P}\overline{Q}x  \ \ \ \ \ \ \  \overline{Q}x=Q\overline{P}x
\end{equation*}
which corrsponds to the code 0122

Case IV:
\begin{equation*}
Px=PQx  \ \ \ \ \ \ \  Qx= \overline{Q}Px 
\end{equation*}
\begin{equation*}
\overline{P}x=P\overline{Q}x  \ \ \ \ \ \ \  \overline{Q}x=\overline{Q}\overline{P}x
\end{equation*}
which corresponds to the code 0221

Case V:
\begin{equation*}
Px=\overline{P}Qx  \ \ \ \ \ \ \  Qx= QPx 
\end{equation*}
\begin{equation*}
\overline{P}x=\overline{P}\overline{Q}x  \ \ \ \ \ \ \  \overline{Q}x=Q\overline{P}x.
\end{equation*}
which corresponds to the code 2102
Q.E.D.


\begin{thebibliography}{999}
\bibitem{Ghazali} M. al-Ghazali, The Incoherence of the Philosophers, Translated by M. E. Marmura, 2nd Edition, Brigham Young University, 2002. 
\bibitem{Birkhoff} G. Birkhoff, Lattice Theory, American Mathematical Society Colloquium Publications, 3rd ed, 1967.
\bibitem{Bombelli1}L. Bombelli, J.-H. Lee, D. Meyer, and R. Sorkin, Spacetime as a causal set, Phys. Rev. Lett. 59, 1987, 521–524.
\bibitem{Bombelli2}L. Bombelli and D. A. Meyer, The origin of Lorentzian geometry, Phys. Lett. A 141,1989, 226–228.
\bibitem{C1} C. Cafaro, W. M. Lord, J. Sun, and E. M. Bollt, Causation Entropy from Symbolic Representations of Dynamical Systems, Chaos 25, 043106, 2015.
\bibitem{C2} C. Cafaro, S. A. Ali, and A. Giffin, Thermodynamic Aspects of Information Transfer in Complex Dynamical Systems, Phys. Rev. E93, 022114, 2016.
\bibitem{C3} C. Cafaro and S. A. Ali, The Spacetime Algebra Approach to Massive Classical Electrodynamics with Magnetic Monopoles, Adv. Appl. Clifford Algebras 17, 23, 2007.
\bibitem{C4}C. Cafaro, Finite-Range Electromagnetic Interaction and Magnetic Charges: Spacetime Algebra or Algebra of Physical Space?, Adv. Appl. Clifford Algebras 17, 617, 2007.
\bibitem{DP} B. A. Davey and H. A. Priestley, Introduction to Lattices and Order, Cambridge University Press, 2002.
\bibitem{DL} C. Doran and A. Lasenby, Geometric Algebra for Physicists, Cambridge University Press, 2003.
\bibitem{Einstein} A. Einstein, On the Electrodynamics of Moving Bodies, Annalen der Physik, 17:891, 1905, Trans. by J. B. Jeffery, 1923.
\bibitem{D} I. Dukovski, Causal structure of Spacetime and Geometric Algebra for Quantum Gravity, Phys. Rev. D87, 064022, 2013.
\bibitem{Euclid} Euclid, The Thirteen Books of The Elements, T. Heath, 2nd ed., Dover, New York, 1956.
\bibitem{GV} M. Grothus and V. Vilasini, Characterizing signalling: Connections Between Causal Inference and Space-time Geometry, (arXiv:gr-qc/2403.0091), 2024.
\bibitem{Fayyad} U. Fayyad, G. Piatetsky-Shapiro, P. Smyth, From Data Mining to Knowledge Discovery in Databases, AI Magazine, 17(3), 37, 1996.
\bibitem{LB} M. Futch, Leibniz's Metaphysics of Time and Space, Springer Science + Business
Media, USA, 2008.
\bibitem{GB} G. Gr$\ddot{a}$tzer, General Lattice Theory, Birkh$\ddot{a}$user Basel, 2003.
\bibitem{Hawking}  S. W. Hawking, A. R. King, and P. J. McCarthy, A New Topology for Curved Space-Time which Incorporates the Causal, Differential and Causal Structures, J. Math. Phys. 17, 1976, no. 2, 174181.
\bibitem{H} D. Hestenes, Spacetime Algebra, Gordon and Breach, New York, 1966.
\bibitem{Jammer} M. Jammer, Concepts of Space: The History of Theories of Space in Physics, 2nd ed., Harvard University Press, Cambridge, Massachusetts, 1969.
\bibitem{KB1} K. H. Knuth, N. Bahreyni, A Potential Foundation for Emergent Space-Time. J. Math. Phys. 55, 112501, 2014; http://dx.doi.org/10.1063/1.4899081. (arXiv:1209.0881[math-ph]).
\bibitem{KB2} K.H. Knuth, N. Bahreyni, The order-theoretic origin of special relativity. P. Bessiere, J.-F. Bercher, A. Mohammad-Djafari (eds.) Bayesian Inference and Maximum Entropy Methods in Science and Engineering, Chamonix, France, AIP Conference Proceedings, American Institute of Physics, Melville NY, 1305, 115–121, 2011.
\bibitem{Kquant} K. H. Knuth, Inferences about interactions: Fermions and the Dirac equation, 2012, (arXiv:1212.2332 [quant-ph]).
\bibitem{Kcontemp} K. H. Knuth, Information-based physics: an observer-centric foundation. Contemporary Physics, 2013, doi:10.1080/00107514.2013.853426. (arXiv:1310.1667 [quant-ph]).
\bibitem{Kelectron} K. H. Knuth, Understanding the Electron, 2015, (arXiv 1511.07766 [physics.gen-ph]).
\bibitem{KBook} K. H. Knuth, Partially Ordered Set. Enumerative Combinatorics, Volume 1, 2nd ed., Cambridge University Press, 2011, pp. 241-463.
\bibitem{Klattice1} K. H. Knuth, Lattice duality: The origin of probability and entropy, Neurocomputing 67C, 2005, 245–274.
\bibitem{Klattice2} K. H. Knuth, Valuations on lattices and their application to information theory, Fuzzy Systems, 2006 IEEE International Conference on, 2006, pp. 217–224.
\bibitem{Klattice3} K. H. Knuth, Measuring on lattices, Bayesian Inference and Maximum Entropy Methods in Science and Engineering, Oxford, MS, USA, 2009 (P. Goggans and C.-Y. Chan, eds.), AIP Conf. Proc. 1193, AIP, New York, 2009, (arXiv:0909.3684v1 [math.GM]), pp. 132–144.
\bibitem{KInfo1}K. H. Knuth, Information physics: The new frontier, Bayesian Inference and Maximum Entropy Methods in Science and Engineering, Chamonix, France, (P. Bessiere, J.-F. Bercher, and A. Mohammad-Djafari, eds.), AIP Conf. Proc. 1305, AIP, New York, 2010, (arXiv:1009.5161v1 [math-ph]), pp. 3–19.
\bibitem{KInfo2}K. H. Knuth, Information-based physics and the influence network, “It from Bit or Bit from It?” FQXi 2013, (arXiv:1308.3337 [quant-ph]).
\bibitem{KS} K. H. Knuth and J. Skilling, Foundations of inference, Axioms 1, 2012, no. 1, 38–73.
\bibitem{KW} K. H. Knuth and J. L. Walsh, An Introduction to Influence Theory: Kinematics and Dynamics, Annalen der Physik, 1700370, 2018, (arXiv:1803.09618 [physics.gen-ph]).
\bibitem{Malament} D. B. Malament, The Class of Continuous Timelike Curves Determines the Topology of Spacetime, J. Math. Phys. 18, (1977), 13991404.
\bibitem{MR} D. Micciancio and O. Regev, Lattice-Based Cryptography. In: Bernstein, D.J., Buchmann, J. and Dahmen, E., Eds., Post-Quantum Cryptography, Springer, Berlin, (2009), 147-191.
\bibitem{OB} N. Ormrod and J. Barrett, Quantum influences and event relativity, 2024, (arXiv:quant-ph/2401.18005).
\bibitem{Reid} D. D. Reid, Manifold dimension of a causal set: Tests in conformally flat spacetimes, Physical Review D, 67(2), 024034, 2003.
\bibitem{SMphys} M.S. Smyth MS and J. H. Martin. X Ray Crystallography. Mol Pathol. (2000) Feb;53(1):8-14. doi: 10.1136/mp.53.1.8. PMID: 10884915; PMCID: PMC1186895.
\bibitem{Sorkin1} R. D. Sorkin, Causal sets: discrete gravity, Lectures on Quantum Gravity (Gomberoff A. and D. Marolf, eds.), Springer US, 2005, (arXiv:gr-qc/0309009), pp. 305–327.
\bibitem{Sorkin2} R. D. Sorkin, Geometry from Order: Causal Sets, 2006, Einstein Online Vol. 2, p. 1007.
\bibitem{Su} S. Surya, The Causal Set Approach to Quantum Gravity, Living Reviews in Relativity 22, 5, 2019.
\bibitem{Su2} S. Surya, Evidence for a Phase Transition in 2D Causal Set Quantum Gravity, Classical and Quantum Gravity, vol. 29, no. 13, 2012, 132001.
\bibitem{Su3} S. Surya, Causal Set Topology, Theoretical Computer Science, Volume 405, Issues 1-2, 2008, pp.187-197.



\end{thebibliography}
\end{document}